%% file: _main.tex
\input{_constants}
\arxiv % \review OR \arxiv OR \cameraready

\documentclass[10pt,twocolumn,letterpaper]{article}
\input{cvpr_header}
\begin{document}

%% TITLE
\title{\paperTitle}
\author{\authorBlock}

\twocolumn[{
\maketitle
\begin{center}
    \captionsetup{type=figure}
    \includegraphics[width=\textwidth]{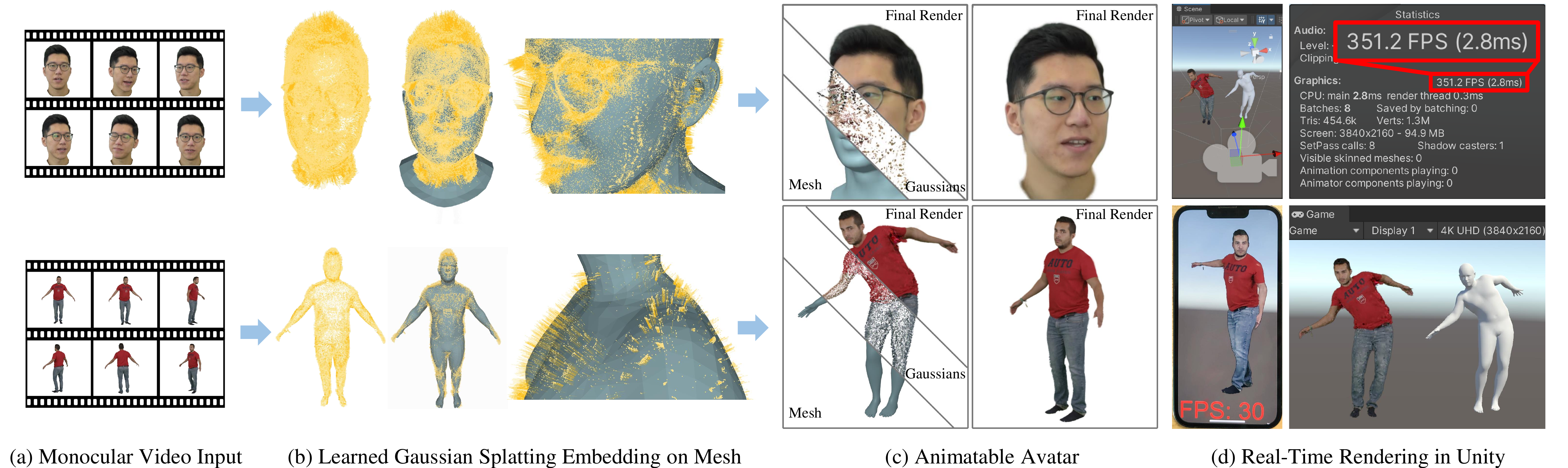}
    \captionof{figure}{
    \textbf{Overview of SplattingAvatar featuring Mesh-Embedded Gaussian Splatting.}
    % SplattingAvatar is a hybrid human avatar modeling framework with Mesh-Embedded Gaussian Splatting, featuring explicit-geometry-implicit-appearance representation. 
    Our method takes (a) monocular videos as input, 
    % while employing a learned Gaussian Splatting embedding technique for mesh placement. 
    while employing (b) a trainable embedding technique for Gaussian-Mesh association.
    % \textbf{(c)} by leveraging the dynamic nature of meshes' animations, the Gaussians can be adjusted to generate realistic rendering results. 
    (c) Animated by mesh through the learned embedding, the Gaussians render into high-fidelity human avatars. 
    (d) SplattingAvatar demonstrates real-time rendering capabilities in Unity, achieving over 300 FPS on an NVIDIA RTX 3090 GPU and 30 FPS on an iPhone 13 (images captured in action).}
    \label{fig:teaser}
\end{center}
}]

\input{00_abstract}

\input{01_intro}

\input{02_related}

\input{03_method}

\input{04_experiments}

\input{10_conclusion}

{\small
\bibliographystyle{ieeenat_fullname}
\bibliography{11_references}
}

\ifarxiv \clearpage \appendix \input{12_appendix} \fi

\end{document}

%% file: _constants.tex
\def\paperID{6146}
\def\confName{CVPR}
\def\confYear{2024}

\def\paperTitle{SplattingAvatar: Realistic Real-Time Human Avatars\\with Mesh-Embedded Gaussian Splatting}

\def\authorBlock{
    Zhijing Shao$^{1,2}$ \qquad
    Zhaolong Wang$^{2}$ \qquad
    Zhuang Li$^{2}$ \qquad
    Duotun Wang$^{1}$ \qquad
    \\
    Xiangru Lin$^{2}$ \qquad
    Yu Zhang$^{2}$ \qquad
    Mingming Fan$^{1,3}$ \qquad
    Zeyu Wang$^{1,3}$ \qquad
    \\
    $^{1}$The Hong Kong University of Science and Technology (Guangzhou)
    \\
    $^{2}$Prometheus Vision Technology Co., Ltd.
    \\
    $^{3}$The Hong Kong University of Science and Technology
}

\newif\ifreview 
\newif\ifarxiv \newcommand{\arxiv}{\arxivtrue}
\newif\ifcamera 
\newif\ifrebuttal 

%% file: cvpr_header.tex
\ifreview \usepackage[review]{cvpr} \fi
\ifarxiv \usepackage[pagenumbers]{cvpr} \fi
\ifrebuttal \usepackage[rebuttal]{cvpr} \fi
\ifcamera \usepackage{cvpr} \fi

\input{_macros}  % Add packages to _macros.tex

\usepackage{xr-hyper}

\makeatletter
\newcommand*{\addFileDependency}[1]{
  \typeout{(#1)}
  \@addtofilelist{#1}
  \IfFileExists{#1}{}{\typeout{No file #1.}}
}

\makeatother

\definecolor{cvprblue}{rgb}{0.21,0.49,0.74}
\usepackage[pagebackref,breaklinks,colorlinks,citecolor=cvprblue]{hyperref}
\usepackage[capitalize]{cleveref}
\crefname{section}{Sec.}{Secs.}
\crefname{table}{Table}{Tables}
\crefname{figure}{Fig.}{Figs.}

\usepackage{ifthen}
\definecolor{MyGreen}{rgb}{0,0.5,0.2}
\definecolor{MyBlue}{rgb}{0,0.2,1.0}
\definecolor{MyRed}{rgb}{1,0.2,0.2}
\definecolor{MyPurple}{rgb}{0.5,0,1.0}
\definecolor{MyOrange}{rgb}{1.0,0.5,0}
\definecolor{MyBrown}{rgb}{0.65,0.35,0}
\definecolor{MyMagenta}{rgb}{1.,0.0,1.}
\definecolor{MyGrey}{rgb}{0.557,0.557,0.557}

\newcommand{\zwc}[1]{{\color{MyRed}{(ZW: #1)}}}
\newcommand{\showcomments}{1}  % comment this line out to hide comments and remove text coloring
\ifthenelse{\isundefined{\showcomments}}
{
\renewcommand{\zwc}[1]{}

}{}
\newcommand{\hidecontents}[1]{}

%% file: _macros.tex
\usepackage{graphicx}	
\usepackage{amsmath}	
\usepackage{amssymb}	
\usepackage{booktabs}
\usepackage{times}
\usepackage{microtype}
\usepackage{epsfig}
\usepackage[table,xcdraw,dvipsnames]{xcolor}
\usepackage{caption}
\usepackage{float}
\usepackage{placeins}
\usepackage{color, colortbl}
\usepackage{stfloats}
\usepackage{enumitem}
\usepackage{tabularx}
\usepackage{xstring}
\usepackage{multirow}
\usepackage{xspace}
\usepackage{url}
\usepackage{subcaption}
\usepackage{xcolor}
\definecolor{redArrow}{RGB}{183,27,20}
\definecolor{greenArrow}{RGB}{117,172,76}
\usepackage[hang,flushmargin]{footmisc}
\usepackage[T1]{fontenc}

\usepackage{algorithm}
\usepackage{algpseudocode}

\ifcamera \usepackage[accsupp]{axessibility} \fi

% ...

\ifarxiv  \fi

\newcommand{\R}[1]{{%
    \textbf{%
        \ifstrequal{#1}{1}{\textcolor{red}{R#1}}{%
        \ifstrequal{#1}{2}{\textcolor{blue}{R#1}}{%
        \ifstrequal{#1}{3}{\textcolor{magenta}{R#1}}{%
        \ifstrequal{#1}{4}{\textcolor{teal}{R#1}}{%
                           \textcolor{cyan}{R#1}%
        }}}}%
    }%
}}

%% file: 00_abstract.tex
\begin{abstract}

We present SplattingAvatar, a hybrid 3D representation of photorealistic human avatars with Gaussian Splatting embedded on a triangle mesh, which renders over 300 FPS on a modern GPU and 30 FPS on a mobile device.
% We disentangle the motion and appearance of a virtual human with explicit mesh geometry and implicit Gaussian Splatting rendering. 
We disentangle the motion and appearance of a virtual human with explicit mesh geometry and implicit appearance modeling with Gaussian Splatting. 
The Gaussians are defined by barycentric coordinates and displacement on a triangle mesh as Phong surfaces. We extend lifted optimization to simultaneously optimize the parameters of the Gaussians while walking on the triangle mesh.
SplattingAvatar is a hybrid representation of virtual humans where the mesh represents low-frequency motion and surface deformation, while the Gaussians take over the high-frequency geometry and detailed appearance.
Unlike existing deformation methods that rely on an MLP-based linear blend skinning (LBS) field for motion, we control the rotation and translation of the Gaussians directly by mesh, which empowers its compatibility with various animation techniques, e.g., skeletal animation, blend shapes, and mesh editing.
% Our method can be trained from monocular videos for full-body or head avatars. We demonstrate state-of-the-art quality and real-time rendering on several datasets. We plan to release our source code in the hope of facilitating research on digital humans.
Trainable from monocular videos for both full-body and head avatars, SplattingAvatar shows state-of-the-art rendering quality across multiple datasets. 
%We plan to release our source code to support further research in digital human modeling.
Code and data are available at \url{https://github.com/initialneil/SplattingAvatar}.
\end{abstract}

%% file: 01_intro.tex
\section{Introduction}
\label{sec:intro}

% To insert a figure: \input{figs/template}
% Or table: \input{tables/template}

% Personalized photorealistic human avatar that is animatable and renders at real-time framerates is widely needed in many downstream applications, e.g., game industry~\cite{zackariasson2012video}, XR storytelling~\cite{ARBook:CE:2014, MRBook:ISMAR:2021}, tele-presentation~\cite{VTour:IEEEVR:2020, VMirror:CHI:2021}, etc.
% Tremendous efforts have been put into making digital humans look real. Practitioners are frequently confronted with a predicament when endeavoring to enhance the quality of their 3D works by augmenting the number of polygons, incorporating multi-layer skin textures~\cite{the_matrix_reload}, implementing advanced hair systems~\cite{Neuralhdhair:CVPR:2022}, and other comparable enhancements, all of which come at the expense of increased computational loads.
The demand for personalized, photorealistic, and animatable human avatars that render in real-time spans a wide array of applications, including gaming~\cite{zackariasson2012video}, extended reality (XR) storytelling~\cite{ARBook:CE:2014, MRBook:ISMAR:2021}, and tele-presentation~\cite{VTour:IEEEVR:2020, VMirror:CHI:2021}. As the quest for digital realism intensifies, practitioners face a growing challenge: improving the quality of 3D human models often means increasing the complexity of these models. This is typically achieved by adding more polygons, layering skin textures~\cite{the_matrix_reload}, and integrating advanced hair systems~\cite{Neuralhdhair:CVPR:2022}. However, these enhancements invariably lead to higher computational demands, creating obstacles in achieving efficiency and portability in avatar rendering.

In our approach, we categorize the representation of mesh-based virtual humans into three distinct levels of detail.
The first two levels encompass body motion and surface deformation, both of which are effectively captured by a mesh~\cite{fvv, neural_head_avatars, xavatar}. 
% This mesh, when combined with accurate texture atlases, inherits a photorealistic quality from real-life images, enabling a vivid representation of dynamic human subjects in a four-dimensional context~\cite{fvv}. 
% Inheriting the photorealistic nature from real images, triangle meshes with proper texture atlas can vividly represent dynamic human subjects in 4D~\cite{fvv}.
The third level, however, focuses on geometric details that are crucial for enhancing realism but challenging to represent with traditional meshes. 
This level is not only computationally demanding to render~\cite{pixel_codec_avatars} but also faces limitations due to the rigid connectivity of mesh vertices, which hinders the adaptability to topological changes and complex or thin structures. 

Recent advances in the field have seen a shift towards using Neural Radiance Fields (NeRF)~\cite{nerf}, especially for capturing high-frequency details in 3D avatar modeling~\cite{neural_actor, neural_body, animatable_nerf, INSTA, MonoAvatar:CVPR:2023, InstantAvatar, mvp, HDHumans}. 
A typical process involves constructing NeRF in a canonical space and then performing volume rendering in the posed space.
This is done by tracing ray samples backward from their posed positions to their canonical origins~\cite{neural_body, animatable_nerf, INSTA, InstantAvatar}. 
However, this reverse mapping process introduces ambiguities, as a single point in the posed space might correspond to multiple points in the canonical space~\cite{snarf, fast-snarf}, leading to challenges in accurately rendering details. 
Additionally, the prevalent use of multilayer perceptron (MLPs) for motion control~\cite{InstantAvatar, xavatar, PointAvatar} tends to overlook the advantages of mesh-based representations for capturing surface deformations, an aspect crucial for realistic avatar movement as highlighted in studies like DECA~\cite{DECA:2021:Siggraph}, CAPE~\cite{CAPE:CVPR:2020}, and TalkSHOW~\cite{talkshow:CVPR:2023}.

To address the challenges posed by the limitations of NeRF and MLP-based motion control in capturing high-frequency details and realistic surface deformations, we introduce a novel solution. 
Inspired by the recently proposed Gaussian Splatting technique~\cite{kerbl3Dgaussians:SIGGRAPH:2023}, we propose explicit motion control of the Gaussians with trainable embeddings on a mesh. The embedding is described by $(k, u, v, d)$ on the mesh as Phong surface~\cite{phongsurface}, where $(u, v)$ represents the local barycentric coordinates of the embedding triangle $k$, and $d$ is the displacement along the interpolated normal vector. 
% The position property of the Gaussians is directly defined by the embedding, while the rotation and scaling are disentangled to pose-dependent and pose-invariant components. 
% The pose-dependent rotation is inferred by the interpolation of per-vertex quaternions which are the weighted average of per-triangle rotations. 
% The pose-dependent scaling adjusts dynamically in response to the area changes of the embedding triangle. 
% Other properties, i.e., canonical rotation and scale, color and opacity, are pose-invariant across frames.
% These pose-dependent modifications allow for a realistic and flexible representation of human motion and surface deformation, while the pose-invariant properties ensure stability and consistency across various poses.
The pose-dependent rotation and scaling adjust dynamically in response to the mesh warping, while the pose-invariant properties, i.e., canonical rotation and scaling, color, and opacity, remain stable and consistent across various poses.
Because the embedding point defined in barycentric coordinates is differentiable only inside the corresponding triangle, cross-triangle updates must be handled properly~\cite{hand:cvpr:2014, hand:siggraph:2016}.
During training, we conduct lifted optimization~\cite{phongsurface} with the embedding points walking on the triangle mesh. 

% We believe that a hybrid representation with Gaussians embedded on triangle mesh will benefit the avatar modeling in three major aspects: 1) the mesh to represent body motion and surface deformation is efficient and highly editable; 2) Gaussian Splatting provides powerful representation for high-frequency geometry and appearance; 3) the embedding allows the Gaussians to be explicitly animated by mesh, resulting in efficient and non-ambiguous motion control with minimal computational load.

% Different from existing hybrid representations~\cite{avatarrex, DELTA:arXiv:2023} that disentangle the avatar modeling by body parts, e.g., separating hair, hands, clothes or face from body, we perform the disentanglement by motion and appearance. In the framework of SplattingAvatar, different parts could have different underlying geometries for motion control but the rendering is uniformly handled by Gaussian Splatting.

Our hybrid representation, Gaussians embedded on a mesh, can be trained from a monocular video and efficiently port to Unity that runs in real time (Figure~\ref{fig:teaser}) by bringing together three key advantages.
% Our belief is that the hybrid representation, combining Gaussians embedded on a triangle mesh, 
First, the use of the mesh for representing body motion and surface deformation not only proves efficient but also allows for high editability. This flexibility is crucial for adapting the avatar to various scenarios and movements. 
Second, the application of Gaussian Splatting enriches this model by providing a robust means to capture high-frequency geometry and appearance details. This is vital for achieving a level of realism that conventional meshes alone cannot offer. 
Third, the embedding technique empowers the Gaussians to be explicitly controlled by the mesh movements. This integration results in an efficient, clear, and non-ambiguous method for motion control, significantly reducing the computational load compared to MLP-based methods.

Furthermore, our approach is distinct from existing hybrid models such as AvatarReX~\cite{avatarrex} and DELTA~\cite{DELTA:arXiv:2023}, which typically segment avatars into body parts like hair, hands, clothes, and face. Instead, our method achieves a disentanglement of motion and appearance. In the SplattingAvatar framework, although different parts may have specific motion control, the rendering is uniformly conducted through Gaussian Splatting. This uniformity achieved by our method ensures a cohesive and harmonious appearance across all parts of the avatar.

\vspace{0.5em}
% We have validated our framework, designed for both head and full-body avatars, across diverse datasets, proving its effectiveness and efficiency. 
%To summarize, we have the following contributions:
% \vspace{-1.0em}
% \hfill \break
\noindent 
\vspace{0.5em}
We summarize our main contributions as follows:
\begin{itemize}[leftmargin=1.25em]
% \vspace{0.5em}
\setlength{\itemsep}{0.25em}
\setlength{\parsep}{0.25em}
\setlength{\parskip}{0.25em}
\item We introduce a framework that integrates Gaussian Splatting with meshes, offering a new avatar representation that achieves realism and computational efficiency.

\item Our approach applies lifted optimization to avatar modeling, allowing for joint optimization of Gaussian parameters and mesh embeddings for accurate reconstruction.

\item We demonstrate the capability of real-time rendering and adaptability to creating diverse avatars through comprehensive evaluation and a Unity implementation.
\end{itemize}
% We demonstrate the effectiveness and efficiency of our framework in datasets for both  head and full-body avatars.
% In summary, our contributions are:

% 1) A novel framework for embedding Gaussian Splatting on triangle meshes, forming a hybrid representation of digital avatars. 

% %2) An extension of lifted optimization for training disentangled explicit-geometry-implicit-appearance avatar models. 
% 2) An extension of lifted optimization for training avatar models by optimizing the parameters of the Gaussians and the trainable embedding simultaneously. 

% 3) An assessment of the realism and applicability of our method across various datasets, coupled with an Unity implementation demonstrating the real-time rendering capabilities and portability of our photorealistic avatar modeling.

% We demonstrate the realism and applicability of our method across different types of datasets, and shows that our photorealistic avatar model can render at real-time frame rate on a mobile device.

%\begin{itemize}
%\item 
%\end{itemize}

%% file: 02_related.tex
\section{Related Work}
\label{sec:related}

\begin{figure*}[htbp]
    \centering
    \includegraphics[width=\linewidth]{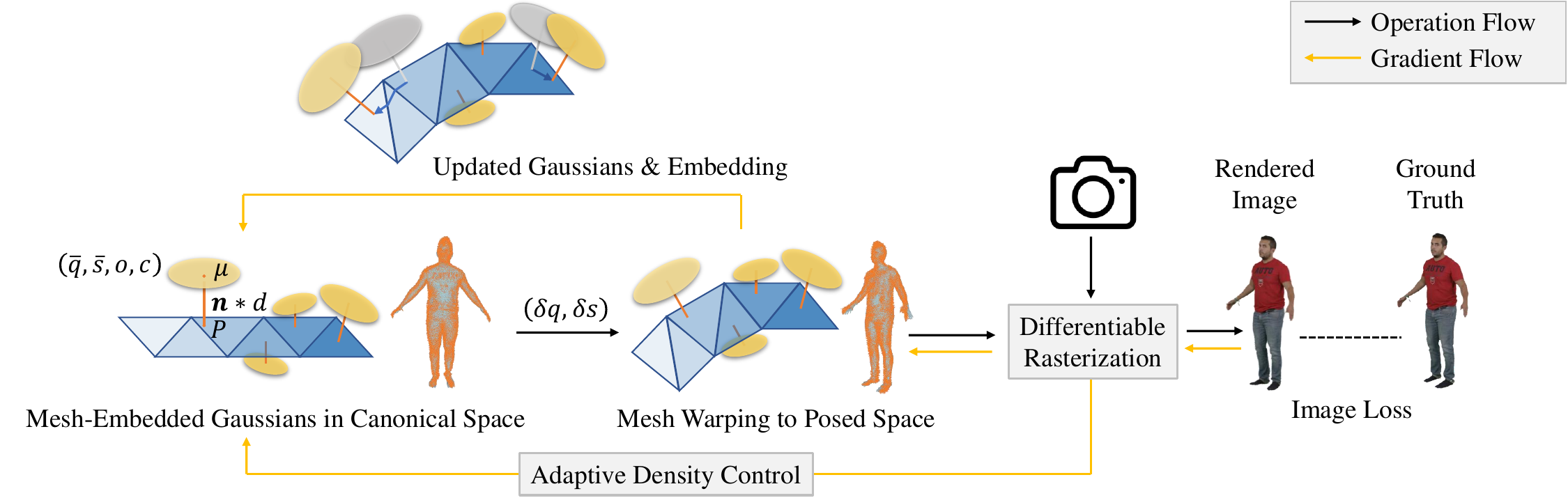}
    \caption{ 
    \textbf{The pipeline of our method.} 
    SplattingAvatar learns 3D Gaussians with trainable embedding on the canonical mesh. 
    The motion and deformation of the mesh explicitly bring the Gaussians to the posed space for differentiable rasterization. Both the Gaussians and embedding parameters are optimized during training. 
    The position ${\mu}$ is the barycentric point $P$ plus a displacement $d$ along the interpolated normal vector $\boldsymbol{n}$. 
    Pose-dependent quaternion and scaling $(\delta{{q}}, \delta{s})$ and pose-invariant quaternion, scaling, opacity, and color $(\overline{{q}}, \overline{s}, o, c)$ together define the properties of the Gaussians. }
    \label{fig:overview}
\end{figure*}

\textbf{Mesh-based avatar.}
The rise of free-viewpoint video in sequences of textured meshes has shown the expressiveness of detailed texture atlas along with as few as 10k triangles~\cite{fvv}. Many efforts~\cite{locally_editable_humans, HDHumans} have been put into extending this line of work to build controllable avatars.
With the help of human shape models with strong prior~\cite{SMPL:TOG:2015, SMPL-X:CVPR:2019, FLAME:SiggraphAsia:2017, REALY:ECCV:2022} that unwrap to a unified UV space, texture atlas can be obtained by 2D image generation supervised through differentiable rendering~\cite{pixel_codec_avatars, clothing_codec_avatar}.
Such prior models provide consistency across large motions and can be recovered from monocular videos or even a single image. To cope with the shape details of identities and clothes, CAPE~\cite{CAPE:CVPR:2020} predicts displacements on the vertices with pose-conditioned VAE.
Due to the limitation of the base model to topological changes, some treat the textured mesh as input conditions~\cite{pixel_codec_avatars, smplpix:WACV:2021} for image rendering, while others resort to implicit representations of the mesh~\cite{snarf, fast-snarf, xavatar, locally_editable_humans}, color~\cite{xavatar, locally_editable_humans, neural_head_avatars}, or materials~\cite{FLARE:SiggraphAsia:2023}.

\noindent\textbf{Implicit neural avatar.}
To achieve convincing rendering beyond the limitation of triangle mesh, especially on the hair, glasses, and clothes, some recent works
~\cite{neural_body, neural_actor, neural_head_avatars, INSTA, MonoAvatar:CVPR:2023, InstantAvatar, monohuman, animatable_nerf, H-NeRF}
focus on constructing NeRF in the canonical space (usually T-pose of SMPL~\cite{SMPL:TOG:2015} or neutral expression of FLAME~\cite{FLAME:SiggraphAsia:2017}) and conduct volume rendering at the posed space. The required backward tracing from pose to canonical is non-trivial and raises an ambiguity issue. Existing works propose to adopt pose conditioned inverse LBS field~\cite{animatable_nerf, vid2avatar:CVPR:2023} or to optimize a root-finding loop with multiple initialization~\cite{snarf, fast-snarf, InstantAvatar}. The increased computational load upon volume rendering prohibits the potential real-time applications.

PointAvatar~\cite{PointAvatar}, with explicit point primitives, takes advantage of forward rasterization that only requires non-ambiguous forward deformation from canonical to pose, producing photo-realistic appearance and detailed challenging geometries such as hair and glasses. In transforming to Gaussian Splatting, we further increase the efficiency and compatibility with our mesh embedding mechanism instead of the LBS-based deformation field and achieve two magnitude faster rendering speed with on-par quality.

\noindent\textbf{Hybrid avatar representation.}
First attempts have been proposed to disentangle human avatar modeling into separate parts with varying properties. AvatarRex~\cite{avatarrex} learns disentangled models for face, body, and hands. SCARF ~\cite{scarf:SiggraphAsia:2022} and DELTA~\cite{DELTA:arXiv:2023} propose hybrid modeling with textured mesh for body, and NeRF for hair and clothing. 
% Another category of hybrid methods focus on combining implicit appearance representation with 
In contrast, our method handles the disentanglement in terms of motion and appearance to explicit mesh geometry and implicit Gaussian Splatting rendering. 
%Multiple mesh parts with different properties/constraints together decide the positions, rotations and scales of the Gaussians that later take over the rendering. 
% The motion and deformation of the mesh decide the positions, rotations and scales of the Gaussians which then take over the rendering.
% The nature of Gaussian Splatting being implicit representation with explicit geometry properties bridges the gap between implicit rendering and explicit motion control.
Different from existing works~\cite{locally_editable_humans, MonoAvatar:CVPR:2023} that attach features to fixed locations on mesh like mesh vertices, our trainable embedding enables the Gaussians to optimize their locations on mesh and distribute unevenly according to the texture complexity.

%% file: 03_method.tex
\section{Method}
\label{sec:method}

\noindent\textbf{Overview.}
Given a sequence of monocular images, each with a registered mesh template, i.e., the deformed mesh of SMPL-X~\cite{SMPL-X:CVPR:2019} or FLAME~\cite{FLAME:SiggraphAsia:2017}, we train a hybrid representation of human avatar as 3D Gaussians~\cite{kerbl3Dgaussians:SIGGRAPH:2023} embedded on the canonical mesh. The Gaussians, parameterized by position, rotation, scale, color, and opacity, are semi-transparent 3D particles that render into camera views through splatting-based rasterization.

Each 3D Gaussian is embedded on one triangle of the canonical mesh in its local $(u, v, d)$ coordinates. The embedding directly defines the position of the Gaussians in both canonical and posed space. Other than position, each Gaussian has its own parameters of rotation, scaling, color, and opacity. With the mesh deformed by animation, the embedding also provides additional rotation and scaling upon each Gaussian. The additional pose-dependent rotation is defined by barycentric interpolated per-vertex quaternion while the additional scaling is defined by the area change of the embedded triangle.

During optimization, the Gaussian parameters and the embedding parameters are updated simultaneously. When the update of $(u, v)$ moves the embedding across the triangle boundary, the barycentric update is re-expressed in the neighboring triangle as if the Gaussian is walking on the mesh. To support embedding, we adapt the clone and split scheme of 3D Gaussians~\cite{kerbl3Dgaussians:SIGGRAPH:2023} to better suit our needs. 

%We tailor the clone and split scheme from~\cite{kerbl3Dgaussians:SIGGRAPH:2023} to support for the embedding.
%and propose to clone nearby Gaussians to unused triangles, encouraging the Gaussians to spread all over the mesh. For Gaussians embedded on the hand area, we pose an additional constraint on $d$ to maintain a reasonable shape of the fingers.

% The original clone and split scheme from \cite{kerbl3Dgaussians:SIGGRAPH:2023} plays an important role in densifying the Gaussians.

\noindent\textbf{Embedding on mesh.}
Inspired by the Phong shading in computer graphics, Phong surface~\cite{phongsurface} defines the position and normal of a point inside a triangle. 
For the point $P$ on triangle $k$ with barycentric coordinate $(u, v)$, its position and normal is a linear interpolation of 
% the triangle's vertices $\{V_{1...3}\}$ and per-vertex normals $\{\boldsymbol{n}_{1...3}\}$:
the triangle's vertices $\{V_1, V_2, V_3\}$ and per-vertex normals 
$\{\boldsymbol{n}_1, \boldsymbol{n}_2, \boldsymbol{n}_3\}$:
\begin{equation}
    %p_{i}(u, v) = u * v1 + v * v2 + (1 - u - v) * v3
    % P_{i}(u, v) = u * V_1 + v * V_2 + (1 - u - v) * V_3
    P = \mathcal{V}(k, u, v) = u * V_1 + v * V_2 + (1 - u - v) * V_3
\end{equation}
\begin{equation}
    %n_{k}(u, v) = u * n1 + v * n2 + (1 - u - v) * n3
    % \boldsymbol{n}_{i}(u, v) = u * \boldsymbol{n}_1 + v * \boldsymbol{n}_2 + (1 - u - v) * \boldsymbol{n}_3
    % \boldsymbol{n}_{i} = \boldsymbol{n}(k, u, v) = u * \boldsymbol{n}_1 + v * \boldsymbol{n}_2 + (1 - u - v) * \boldsymbol{n}_3
    \boldsymbol{n} = \mathcal{N}(k, u, v) = u * \boldsymbol{n}_1 + v * \boldsymbol{n}_2 + (1 - u - v) * \boldsymbol{n}_3
\end{equation}
where $\mathcal{V}$ maps triangle index $k$ and barycentric coordinates $(u, v)$ to a point on the mesh and $\mathcal{N}$ the interpolated normal.

We define the position of a Gaussian, i.e., the mean ${\mu}$, by a displacement $d$ along the interpolated normal vector:
\begin{equation}
    {\mu} = P + d * \boldsymbol{n}
\end{equation}
Embedding $E=\{k, u, v, d\}$ approximates a first-order continuous space around the mesh surface.

As proposed by Zielonka et al.~\cite{INSTA}, for the corresponding triangle in the canonical and posed space at frame $t$ we compute the matrix $\{R_{cano}, R_{pose}\}$ based on the triangle's tangent, bitangent, and normal to track the triangle rotation from canonical to pose, noted that the notation $t$ is skipped. The rotation matrix is then converted to a quaternion, and we calculate the per-vertex quaternion ${q}_{V}$ by area-weighted average from surrounding neighbor triangles:
\begin{equation}
    R_{k} = R_{cano} R_{pose}^{-1}
\end{equation}
\begin{equation}
%\upsilon
    {q}_{V} = \frac
    {\sum_{k \in \Omega_V}{A_k {q}_k}}
    {\sum_{k \in \Omega_V}{A_k}}
\end{equation}
where $\Omega_V$ is the neighbor triangles of vertex $V$, $A_k$ and ${q}_k$ are the triangle's area and quaternion respectively. For an embedding $E_i$ with quaternions $\{{q}_1, {q}_2, {q}_3 \}$ calculated on the corresponding triangle vertices at frame $t$, the barycentric interpolated rotation $\delta {q}_{i,t}$ is multiplied to the canonical rotation $\boldsymbol{\overline{q}}_i$ of the Gaussian in the canonical space:
\begin{equation}
    \delta {q}_{i,t} = u * {q}_1 + v * {q}_2 + (1 - u - v) * {q}_3
\end{equation}
\begin{equation}
    {q}_{i,t} = \delta {q}_{i,t} * {\overline{q}}_i 
\end{equation}
The same applies to scaling where the area change of the embedded triangle is used to represent the scaling caused by deformation: $s_{i, t} = ({A_{pose}}/{A_{cano}}) \overline{s}_i$. While the original implementation of Gaussian Splatting~\cite{kerbl3Dgaussians:SIGGRAPH:2023} represents color in view-dependent spherical harmonics, we choose to turn it off for simplicity~\cite{luiten2023dynamic:arxiv:2023}.

\begin{figure}
    \centering
    \includegraphics[width=\linewidth]{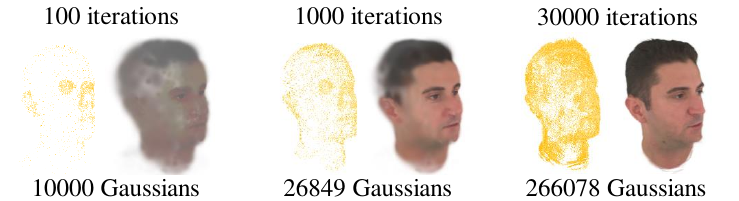}
    \caption{
        \textbf{The development of Gaussian embeddings on mesh.}
        %The red dots are the embedding points on mesh. The blue dots are the means of the Gaussians and the white lines depicts the displacements along the normal vector. 
        Each line segment indicates the position of one Gaussian displaced from its embedding point on mesh.
        Gaussians for off-surface geometries like the hair have positive displacements while others turn to have negative displacements because when the mesh surface is correctly aligned to the geometry like in the facial area, the means for the Gaussians will be inside the mesh.
        % a) Initial embeddings are randomly selected on mesh surface. b) Movement of the Gaussians cause erroneous early embeddings. c) Finally the embeddings end up evenly spreading the mesh with reasonable displacements.
    }
    \label{fig:embedding}
\end{figure}

We perform initialization by randomly selecting 10k pairs of triangle indices and barycentric coordinates on the canonical mesh. We set the barycentric coordinates to be the current $(u, v)$ of embeddings and initialize all the $d$ to be zero. With the position of the Gaussians calculated from the embeddings, we initialize other properties of the Gaussians according to their original definitions~\cite{kerbl3Dgaussians:SIGGRAPH:2023}.
%We observe more erroneous embedding in the beginning of the training, e.g. in the rest pose when arms are close to body, Gaussians from the body might move to cover for arms. With the training proceed with more poses, inappropriate embeddings that drive the Gaussians to empty space or to conflict with others, will be pruned. More correct embeddings will emerge in the clone and splitting process. 
Initially, the Gaussians are positioned on the surface of the mesh. With the training proceeds with more poses, the embeddings generally bring the Gaussians to approximate the actual geometry and densify in the regions with rich texture.
Figure \ref{fig:embedding} illustrates the development of the embeddings.

% \textbf{to-do:} discuss constraint on hands. 
% Due to inaccurate mesh registration in the training data, a naive implementation of adding loss on $d$ would resulting Gaussians being removed.

\noindent\textbf{Differentiable rendering of Gaussian Splatting.}
With the position, rotation, and scaling of the Gaussians updated by the mesh deformation at frame $t$, we perform differentiable Gaussian rendering~\cite{kerbl3Dgaussians:SIGGRAPH:2023} to the observed camera view(s). 
The Gaussian in space is defined by its mean ${\mu}$ and a 3D covariance matrix $\Sigma$.
\begin{equation}
    G_{i,t}(x) = e^{- \frac{1}{2} (x)^T \Sigma_{i,t}^{-1} (x)}
\end{equation}
\begin{equation}
    \Sigma_{i,t} = R_{i,t} S_{i,t} S_{i,t}^T R_{i,t}^T
\end{equation}
where $R_{i,t}$ is the rotation matrix constructed from ${q}_{i,t}$, and $S_{i,t}$ the scaling matrix from $s_{i,t}$. 
Given the world-to-camera view matrix $W$ and the Jacobian $J$ of the point projection matrix. The influence of the Gaussian is splatted to 2D~\cite{SurfaceSplatting}:
\begin{equation}
    \Sigma' = J W \Sigma W^T J^T
\end{equation}
The image formation of Gaussian Splatting is akin to NeRF, where the same volume rendering formula is applied to the blending from near to far. 
The color $C$ of a pixel rendered by $N$ Gaussians is given by a series of \emph{$\alpha$-blending}:
\begin{equation}
    C = \sum_{i=1}^{N} {c_i} \alpha_i \prod_{j=1}^{i-1}(1 - \alpha_j)
    \label{blending}
\end{equation}
with $\alpha_i$ evaluated from the 2D covariance, and an opacity in logit $o_i$ with sigm() being the standard sigmoid function:
\begin{equation}
    \alpha_i(P) = \text{sigm}(o_i) \exp( -\frac{1}{2} (P - \mu_i) (\Sigma_i)^{-1} (P - \mu_i) )
\end{equation}
% where sigm() is the standard sigmoid function.

The Equation~\ref{blending} is implemented in CUDA with a for loop for each pixel, while in our Unity implementation, each Gaussian is drawn by a front-parallel Quad primitive based on the projection and 2D covariance. We resort to the standard rasterization pipeline of the rendering engine to enable \emph{$\alpha$-blending} with these semi-transparent Gaussians.

% The original Gaussians are omnidirectionally visible. Due to the very limited visual supervision from monocular video, we propose to perform \emph{normal culling} with the normal vector from embedding, so that the Gaussians embedded on back-sided triangles r.w.t. the current camera will not be rendered. 
% We also add a weak sparsity regularization on opacity to remove redundant Gaussians that are never used.
% Due to the very limited visual supervision from monocular video, we propose two regularization terms: a opacity term to remove redundant Gaussians that are never used and a scaling term to prevent Gaussians from growing long and thin.
Due to limited viewing angle and pose variations from monocular video, we propose a scaling regularization term to prevent Gaussians from growing long and thin.
A random background color is generated every iteration to mix with the rendered image $I$ and ground truth image $I_{gt}$, providing important cues for the silhouette. 
The photometric loss is the sum of $\mathcal{L}_1$ with perceptual loss~\cite{zhang2018perceptual}.

\begin{equation}
    \mathcal{L} = \mathcal{L}_1 + \lambda_{l} \mathcal{L}_{lpips} 
    % + \lambda_{o} \mathcal{L}_{opacity} 
    + \lambda_{s} \mathcal{L}_{scaling}
\end{equation}
% \begin{equation}
%     \mathcal{L}_{opacity} = \frac{1}{N} \sum_{i=1}^{N}{ \vert sigm(o_i) \vert }
% \end{equation}
\begin{equation}
\mathcal{L}_{scaling} (i) = 
\left\{
     \begin{array}{lr}
     \vert \hat{s_i} \vert, &  \hat{s_i} > \max(T_s, T_r \check{s_i}) \\
     0, & otherwise \\
     \end{array}
\right.
\end{equation}

With $s_i \in \mathbb{R}^3$ being the scaling of a Gaussian, $\hat{s_i}$ and $\check{s_i}$ are the maximum and minimum scaling values respectively. 
The scaling regularization is posed on $\hat{s_i}$ when it is both long (larger than $T_s$) and thin (larger than $T_r$ times $\check{s_i}$).
Please see Section~\ref{subsec:ablation} for an ablation study on the regularization term.
%We use $\lambda_l = 0.01$, $\lambda_s = 1.0$, $T_s = 10.0$ and $T_r = 0.008$ all through the experiments.

\begin{figure}
    \centering
    \includegraphics[width=\linewidth]{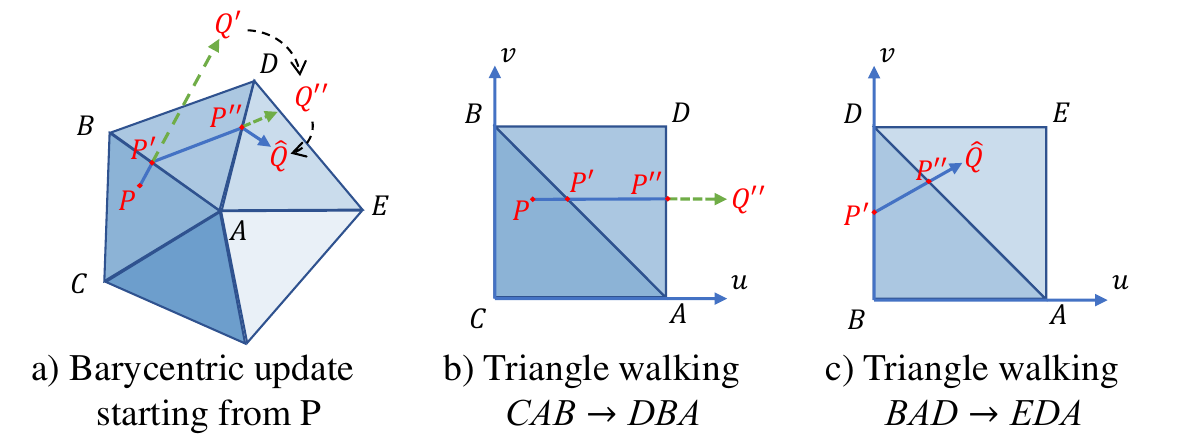}
    \caption{
    \textbf{Walking on triangles for embedding update.} 
    a) The recursion process of walking on a triangle mesh. 
    b) The update $P + \delta$ starting from triangle \emph{CAB} is re-expressed as $P' + \delta'$ in triangle \emph{DBA}, and c) re-expressed again in \emph{EDA}. 
    %The barycentric coordinates is re-expressed by treating two right triangles adjacent to each other on the hypotenuse. 
    The re-expression between two triangles is conducted by conceptually treating them as two right triangles adjacent to each other on the hypotenuse.
    }
    \label{fig:triangle}
\end{figure}

\noindent\textbf{Walking on a triangle mesh.}
The notion \emph{Lifted Optimization} arises in the model-point registration for hand tracking~\cite{hand:cvpr:2014, hand:siggraph:2016, phongsurface} in contrast to \emph{Iterative Closest Point (ICP)}, where the solve for model pose and correspondences are \emph{lifted} to be \emph{simultaneous}. 
We extend this notion to our avatar training, %where the Gaussians strive to move to correct poses in space while their correspondences to mesh, i.e. embedding, update simultaneously.
where the properties of the Gaussians and the trainable embeddings are optimized simultaneously.
The barycentric coordinate of a point $P$ is $(k, u, v)$ defined within triangle $k$. When the learned update $Q = (k, u, v) + (\delta u, \delta v)$ is outside triangle $k$, we find the intersection $P'$ on the shared edge of the adjacent triangle $k'$ and re-express the remaining update in $k'$ as $Q' = P' + (\delta u', \delta v')$. Because the barycentric coordinates are agnostic to the triangle shape, without loss of generality, the re-expression is conducted by conceptually treating two adjacent triangles as right triangles with the intersection on the hypotenuse. 
The update is iteratively re-expressed until it ends inside the final triangle. 
We show the re-expression process in Figure~\ref{fig:triangle}.
The detailed steps are presented in Algorithm~\ref{triangle_walk}. Noted that we omit the conceptual re-ordering of the vertices.

\begin{algorithm}
    \caption{Walking on triangles}
    
    \textbf{Input: $k, u, v, \delta u, \delta v$}
    
    \textbf{Output: $\hat{k}, \hat{u}, \hat{v}$}

    \begin{algorithmic}
    \Function{WalkOnTriangles}{$k, u, v, \delta u, \delta v$}
        \State $P \gets (u, v)$
        \State $Q \gets (u + \delta u, v + \delta v)$
        \If {Q is inside triangle}
            \State Return $(k, Q.u, Q.v)$ 
        \EndIf

        \State Intersect \emph{P-Q} with hypotenuse* on $(u', v')$
        \State                               \Comment{*reorder vertices if needed}
        \State {$\delta{u'} \gets \delta{u} - (u' - u)$}
        \State {$\delta{v'} \gets \delta{v} - (v' - v)$}
        \State Return ReExpress$(k, u', v', \delta{u'}, \delta{v'})$
    \EndFunction

    \Function{ReExpress}{$k, u', v', \delta{u'}, \delta{v'}$}
        \State $\hat{k} \gets$ adjacent of $k$
        \State $\hat{u} \gets 1 - u'$, $\hat{v} \gets 1 - v'$
        \State $\delta \hat{u} \gets - \delta u'$, $\delta \hat{v} \gets - \delta v'$
        \State Return WalkOnTriangles$(\hat{k}, \hat{u}, \hat{v}, \delta \hat{u}, \delta \hat{v})$
    \EndFunction
    
    \end{algorithmic}

    {$(\hat{k}, \hat{u}, \hat{v}) \gets$ WalkOnTriangles$(k, u, v, \delta u, \delta v)$}

    \label{triangle_walk}
\end{algorithm}

\noindent\textbf{Optimization.}
We use Adam to optimize the Gaussian parameters and the embedding parameters. The original learning rate attenuation on position~\cite{kerbl3Dgaussians:SIGGRAPH:2023} is instead applied to the embedding parameters. We record the current barycentric $(u, v)$ and optimize for $(\delta{u}, \delta{v}, d)$. 
%On every 100 iterations we apply triangle walk with the accumulated $(\delta{u}, \delta{v})$. 
The triangle walking in Algorithm~\ref{triangle_walk} is implemented as a \emph{pybind11} module in C++.
When an embedding is being transferred to another triangle, we reset its corresponding optimizer state of the $(\delta{u}, \delta{v}, d)$.

%  ~\cite{pybind11}

The densification process~\cite{kerbl3Dgaussians:SIGGRAPH:2023} plays an important role in allocating more Gaussians to where in need. In the clone and prune process, the embedding parameters are copied or deleted in the same way as Gaussian parameters. In the split process, when a new position $\hat{{\mu}}$ is sampled from the Gaussian, we solve a mini problem with triangle walking to find the new embedding:
\begin{equation}
     %text{argmin}
    \hat{E} = \underset{k, u, v, d}{\arg\min} \Vert \mathcal{V}(k, u, v) + d * \mathcal{N}(k, u, v) - \hat{{\mu}} \Vert_{2}^{2}
\end{equation}
%To help convergence, we propose to add a new densification scheme. We list all triangles without embedding and clone the nearest Gaussian to the center of the triangle. We find this strategy to provide meaningful and efficient birth place of new Gaussians and helps convergence.
% We follow XXXXX~\cite{kerbl3Dgaussians:SIGGRAPH:2023} to reset the opacity of all Gaussians occasionally. 
% The opacity values of all Gaussians are reset to zero occasionally~\cite{kerbl3Dgaussians:SIGGRAPH:2023}.
% We find this to be effective in terms of removing redundant Gaussians.

% We find that casual poses are way more often than rare poses in the training footage. To prevent over-fitting to casual poses, we ...\emph{to-to}
% To prevent over-fitting to casual poses, we record the rendering error $E_{t, i}$ (psnr???) on every view at every frame, and set the frame error to be the maximum of all views: $E_{t} = \max_{i}\{E_{t, i}\}$. On every training batch, we apply importance sampling based on the frame error for frame selection. Additionally for multiview dataset, we randomly select 1/4 of the total views (or at least 4 views) for a faster transition between different poses.

\noindent\textbf{Unity implementation for mobile device.}
With maximum compatibility in mind, we made SplattingAvatar solely rely on the warped mesh. Before exporting to Unity, we uploaded the canonical mesh in \emph{.obj} format to Mixamo~\cite{Mixamo} for auto-rigging. In total, we exported one \emph{.ply} file of the Gaussians, one \emph{.json} file describing the embedding, and one \emph{.fbx} file from Mixamo to Unity. Note that the \emph{.fbx} file can be rigged by any other software for customized needs, as long as the triangle order is maintained.

We implemented the Gaussian renderer in Unity's compute shaders, starting from sorting all Gaussians by the z-axis in camera coordinates from near to far. Based on the calculated 2D covariance $\Sigma'$, one front-parallel quad primitive is drawn for every visible Gaussian centered at its position. 
This one-primitive-one-Gaussian strategy is important for the game engine to properly handle the occlusion of other regular objects.
For every pixel to draw in the fragment shader, our implementation emits color with alpha pre-multiplication and sets the blend function to \emph{(ONE, ONE\_MINUS\_SRC\_ALPHA)}. 
Our Unity program achieves a high performance of over 300 FPS on a modern GPU while maintaining a steady 30 FPS on an iPhone 13.
%, where the compute shader implementation of Radix sort occupies xxx of the total time budget. Please see table xx for the breakdown.

%% file: 04_experiments.tex
% \section{Experimental Results}
\section{Experiments}
\label{sec:experiments}

\begin{figure*}[ht]
    \centering
    \includegraphics[width=\linewidth]{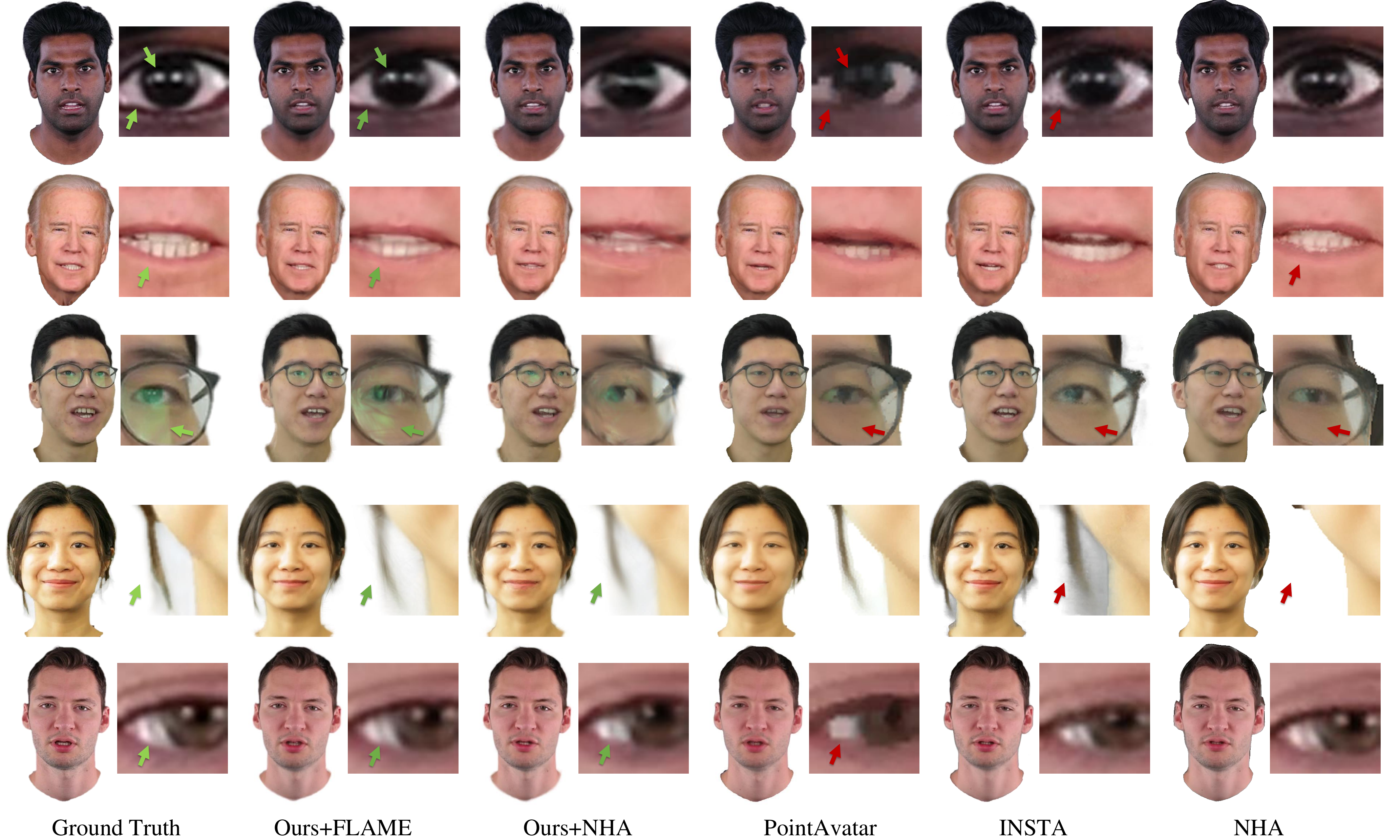}
    \caption{ 
        \textbf{Qualitative comparison on head avatars.} 
        SplattingAvatar produces photorealistic rendering for avatars with high-quality details especially in the eye and hair regions. Even the light reflection on the glasses is well reconstructed. Both PointAvatar~\cite{PointAvatar} and NHA~\cite{neural_head_avatars} can reconstruct good geometries but the rendering quality is limited by their underlying representations, i.e., points and texture atlas respectively. Compared to INSTA~\cite{INSTA}, our trainable embedding scheme produces better quality for off-surface geometries, especially for the glasses. 
        The \textcolor{greenArrow}{green} arrows highlight where our results have better consistency with Ground Truth, while the 
        \textcolor{redArrow}{red} arrows point to where other methods show significant artifacts or noise.
        Please see the supplemental materials for illustrations of the error map.
    }
% \textcolor{ForestGreen}{green} 
% \textcolor{RubineRed}{red}
    \label{fig:exp1}
\end{figure*}

\begin{table}[t]
\centering
% \resizebox{\columnwidth}{!}{%
\begin{tabular}{lccc}
    \toprule
    Method & PSNR$\uparrow$ & SSIM$\uparrow$ & LPIPS$\downarrow$ \\
    \midrule
    NHA~\cite{neural_head_avatars} & 20.29 & 0.883 & 0.145 \\
    INSTA~\cite{INSTA} & 26.42 & 0.924 & 0.080 \\
    PointAvatar~\cite{PointAvatar} & 27.84 & 0.913 & 0.067 \\
    \midrule
    Ours+FLAME & \underline{28.19} & \textbf{0.931} & \underline{0.063} \\
    Ours+NHA & \textbf{28.86} & \textbf{0.931} & \textbf{0.060} \\
    \bottomrule
\end{tabular}
% }
\caption{
    \textbf{Quantitative comparison on head avatars.}
    Both variations of our method outperform existing methods in terms of average photometric errors. 
    With detailed meshes from NHA~\cite{neural_head_avatars}, \emph{Ours+NHA} performs the best based on the metrics. However, we observe better visual quality with \emph{Ours+FLAME} in the inner regions of the rendered image. 
    %However, \emph{Ours+FLAME} achieves the lowest perceptual error because the mesh of FLAME contains less triangle distortions.
}
\vspace{-0.05in}
\label{tab:exp1}
\end{table}

\begin{figure*}[ht]
    \centering
    \includegraphics[width=\linewidth]{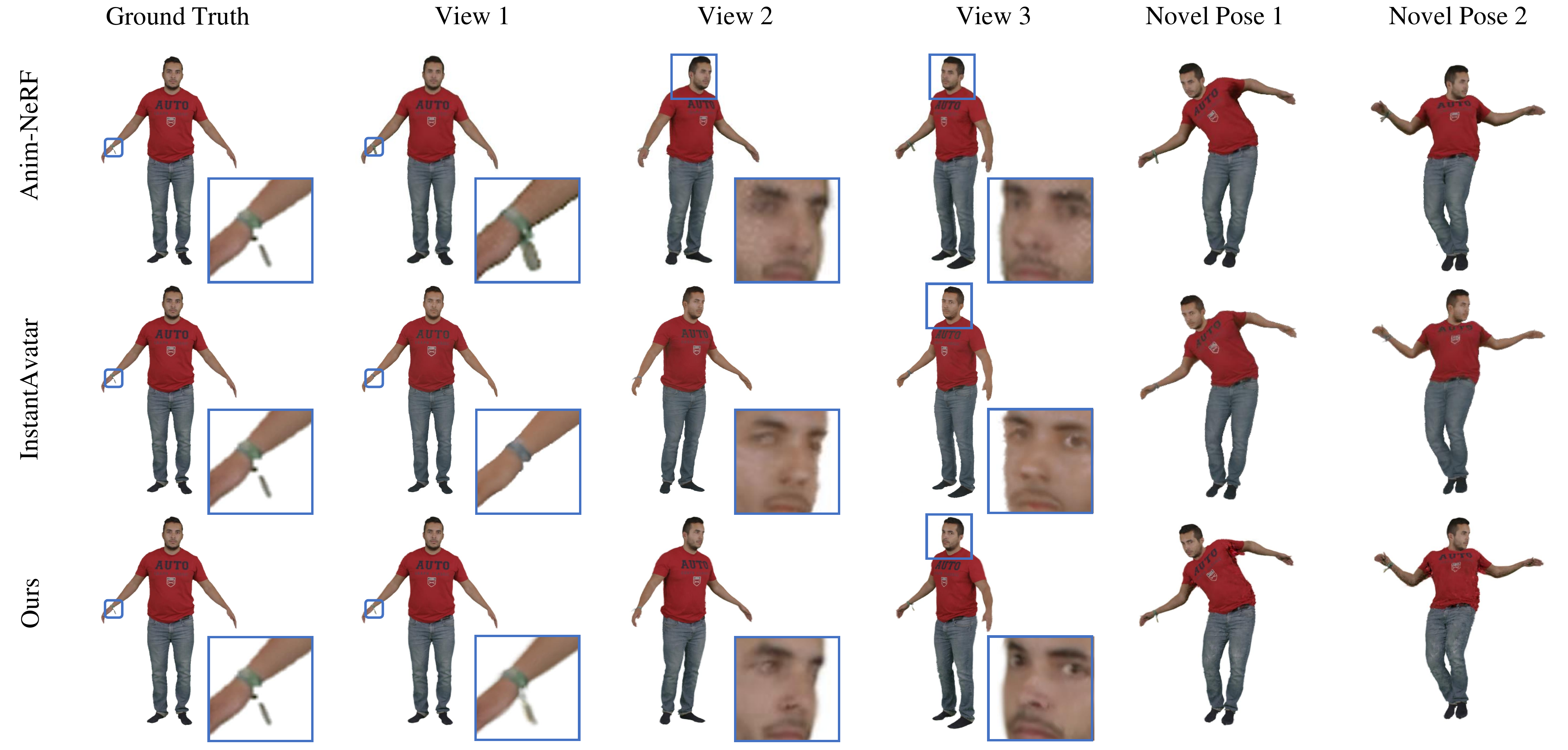}
    \caption{ 
        \textbf{Qualitative comparison on PeopleSnapshot~\cite{peoplesnapshot:3DV:2018}.} 
        We show the results on PeopleSnapshot (columns 2--4) and novel pose animation (columns 5--6). SplattingAvatar produces photorealistic rendering for full-body avatars, especially in the facial area, and captures thin structures like the accessory on the wrist. 
    }
    \label{fig:exp_people}
\end{figure*}

\begin{table*}[t]
\centering
\resizebox{\textwidth}{!}{
\begin{tabular}{lcccccccccccc}
    \toprule
    & \multicolumn{3}{c}{male-3-casual} & \multicolumn{3}{c}{male-4-casual} & \multicolumn{3}{c}{female-3-casual} & \multicolumn{3}{c}{female-4-casual}  \\
    & PSNR$\uparrow$ & SSIM$\uparrow$ & LPIPS$\downarrow$& PSNR$\uparrow$ & SSIM$\uparrow$ & LPIPS$\downarrow$& PSNR$\uparrow$ & SSIM$\uparrow$ & LPIPS$\downarrow$& PSNR$\uparrow$ & SSIM$\uparrow$ & LPIPS$\downarrow$ \\
    \midrule
    Anim-NeRF~\cite{Anim-NeRF} & 29.37 & 0.970 & \textbf{0.017} & 28.37 & 0.960 & \underline{0.027} & 28.91 & 0.974 & \textbf{0.022} & 28.90 & 0.968 & \textbf{0.017} \\
    InstantAvatar~\cite{InstantAvatar} & 30.91 & 0.977 & 0.022 & 29.77 & 0.980 & \textbf{0.025} & 29.73 & 0.975 & \underline{0.025} & 30.92 & 0.977 & 0.021 \\
    \midrule
    Ours & \textbf{33.01} & \textbf{0.982} & \underline{0.020} & \textbf{30.99} & \textbf{0.982} & 0.029 & \textbf{30.81} & \textbf{0.978} & 0.028 & \textbf{32.57} & \textbf{0.981} & \underline{0.018} \\
    \bottomrule
\end{tabular}
}
\caption{
    \textbf{Quantitative comparison on PeopleSnapshot.}
    Compared to two SoTA methods, we achieve significant improvements in pixel-wise quality with PSNR and SSIM. 
    All three methods achieve good perceptual quality in terms of LPIPS where the metrics are close.
}
\vspace{-0.05in}
\label{tab:exp_people}
\end{table*}

To demonstrate the effectiveness of SplattingAvatar, we compared it with state-of-the-art (SoTA) methods in two different types of datasets for head and full-body avatars.

\subsection{Datasets}
\noindent\textbf{Monocular video for head avatar.}
Taking a single monocular video to construct a head avatar for the given subject, our method takes as input images, masks, camera parameters, and tracked FLAME meshes, denoting \textbf{\emph{Ours+FLAME}}. 
We evaluated our approach with several SoTA methods on a combined dataset from NHA~\cite{neural_head_avatars}, NerFace~\cite{NerFace:CVPR:2021}, INSTA~\cite{INSTA} and PointAvatar~\cite{PointAvatar}, including 10 subjects covering different videos captured with DSLR, smartphones and from the Internet. The pre-processing pipeline of IMavatar~\cite{IMavatar:CVPR:2022} and INSTA~\cite{INSTA} was altered to apply DECA~\cite{DECA:2021:Siggraph} for face tracking, RVM~\cite{RVM:WACV:2021} for segmentation, and BisenetV2~\cite{BisenetV2} for face parsing. For each video, the last 350 frames were used as testing samples. Because our method can directly be animated by the given mesh, we further unleashed its potential by training and testing on the generated meshes from NHA~\cite{neural_head_avatars} which have more geometry details. This variation is referred to as \emph{Ours+NHA}.   %really need the BF?  \textrbf{}

%\vspace{-1mm}
\noindent\textbf{PeopleSnapshot.}
We conducted a quantitative evaluation of the rendering quality of full-body avatars on the PeopleSnapshot~\cite{peoplesnapshot:3DV:2018} dataset, which captures the human subjects rotating in A-pose. Following the protocol of InstantAvatar~\cite{InstantAvatar}, we used SMPL meshes refined by Anim-NeRF~\cite{Anim-NeRF}. Our method demonstrates the generalizability to novel poses through qualitative analysis in Section~\ref{subsec:comparison}.

% {
% \setlength{\parindent}{0cm}
% % \subsection{Monocular video for full-body avatar}
% % \textbf{Monocular video for full-body avatar.}
% \textbf{SHOW-4sub}
% }
% SHOW~\cite{talkshow:CVPR:2023} is a dataset of monocular videos of talk show hosts with rich facial expressions and hand gestures.
% The pseudo ground truth of full-body pose estimation is generated by adapting SMPLify-X~\cite{SMPL-X:CVPR:2019} with multiple state-of-the-art algorithms, including PIXIE~\cite{PIXIE:3DV:2021}, PyMAF-X~\cite{pymafx2023} and DECA~\cite{DECA:2021:Siggraph} for initialization, and OpenPose~\cite{openpose:cvpr:2023}, DeepLab V3, MediaPipe~\cite{mediapipe} and MICA~\cite{MICA:ECCV2022} for joint optimization.
% We process 4 subjects from the SHOW dataset, denoting \emph{SHOW-4sub}, each with 2000-2400 frames for training and 600 frames for testing.

\subsection{Comparison with SoTA} \label{subsec:comparison}
\noindent \textbf{Head avatar.}
To evaluate the rendering quality of the learned avatars, we 
animated SplattingAvatar with the registered meshes of testing images. For \emph{Ours+NHA}, we trained NHA~\cite{neural_head_avatars} on the training set and extracted the final meshes for both the training and testing images, which were further used for the training and testing of our method respectively.

We conducted a comparative analysis of SplattingAvatar against INSTA~\cite{INSTA}, PointAvatar~\cite{PointAvatar}, and NHA ~\cite{neural_head_avatars}. As depicted in Figure~\ref{fig:exp1}, our method achieves superior quality in terms of improved details in the eye and hair regions, and even being able to capture the light reflection on the glasses. For \emph{Ours+FLAME}, though the off-surface geometries like hair and glasses are not fully represented by meshes, our method can handle the rendering decently because the embeddings are optimized to find correct motions from nearby triangles.
Please see Table~\ref{tab:exp1} for quantitative evaluations with PSNR, SSIM and LPIPS.
% Please see the supplemental materials for more detailed comparisons.

%where PointAvatar \cite{PointAvatar} is based on point-NeRF to capture explicit geometry property and INSTA \cite{INSTA} is based on Instant-NGP \cite{instant} to achieve fast training and high quality rendering.

\noindent\textbf{Full-body avatar.}
We made a comparison to InstantAvatar~\cite{InstantAvatar} and Anim-NeRF~\cite{Anim-NeRF} on PeopleSnapshot. For InstantAvatar~\cite{InstantAvatar}, a complete training was performed for 200 epochs as suggested in the most recent version of the author's code. Image quality metrics in Table~\ref{tab:exp_people} demonstrate the effectiveness of our method in terms of the lowest pixel-wise errors. We qualitatively show the comparison of rendering quality of testing images in Figure~\ref{fig:exp_people}, together with the demonstration of generalizability to novel poses. Our representation is friendly to thin structures like the accessory on the wrist.
% For novel poses, We produce better quality in the facial area comparing to InstantAvatar~\cite{InstantAvatar}. But due to very limited pose variation in the training set, our results undergo slightly more artifacts under the shoulder areas. 
% For novel poses, 
Our approach produced better quality overall and especially in the facial area compared to InstantAvatar~\cite{InstantAvatar}, but slightly more artifacts under the shoulder due to very limited pose variations in the training set. We believe this can be much improved with more training poses.
%We further compare to InstantAvatar~\cite{InstantAvatar} in SHOW-4sub.

% \begin{table}[t]
% \centering
% % \resizebox{\columnwidth}{!}{%
% \begin{tabular}{lccc}
%     \toprule
%     Method & PSNR$\uparrow$ & SSIM$\uparrow$ & LPIPS$\downarrow$ \\
%     \midrule
%     Anim-NeRF~\cite{Anim-NeRF} & 28.88 & 0.9682 & 0.0206 \\
%     InstantAvatar~\cite{InstantAvatar} & 30.33 & 0.976 & 0.023 \\
%     \midrule
%     Ours & \textbf{31.88} & \textbf{0.9812} & \textbf{0.0244} \\
%     \bottomrule
% \end{tabular}
% % }
% \caption{
%     \textbf{Quantitative comparison on PeopleSnapshot.}
% }
% \vspace{-0.05in}
% \label{tab:exp_people}
% \end{table}

\subsection{Ablation Study} \label{subsec:ablation}
\begin{figure}[ht]
    \includegraphics[width=\linewidth]{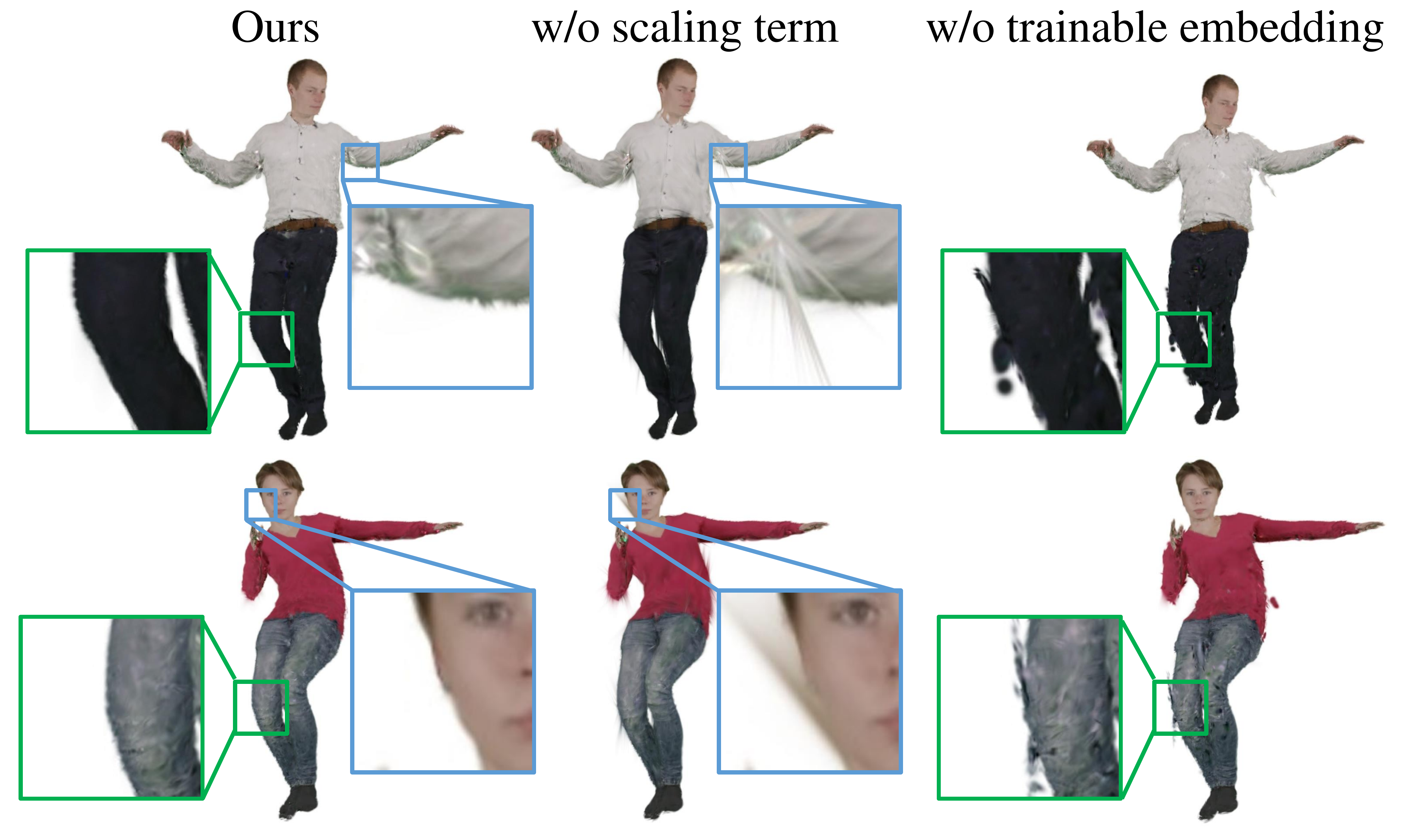}
    \caption{ 
        % \textbf{Effect of scaling regularization.} 
        % Without the scaling regularization term, Gaussians that are long and thin cause visible artifacts. Our scaling term successfully removes most of the artifacts when animated by novel poses.
        \textbf{Ablation study.} 
        Without the scaling regularization term, Gaussians that are long and thin cause needle-like artifacts. Without trainable embedding, Gaussians do not follow the movement of the mesh tightly, leading to irregular rendering results. The application of our trainable embedding and the scaling term successfully removes most of the artifacts when rendered into novel poses. 
    }
    \label{fig:ablation}
\end{figure}

\textbf{Trainable embedding.}
The key component of our method is the trainable embedding on the mesh. We conducted an ablation experiment by replacing it with fixed embedding on mesh and a trainable local shift $\Delta x \in \mathbb{R}^3$ per Gaussian.
% The metrics in Table.~\ref{tab:abalation_nouvd} shows the necessity of the trainable embedding.
Without trainable embedding, the Gaussians encountered difficulties in following the mesh correctly.
The right column of Figure~\ref{fig:ablation} shows the irregular rendering artifacts without trainable embedding.

\noindent \textbf{Regularization.}
% With limited supervision from monocular video, some Gaussians are never used after initialization.
In the optimization process of Gaussian Splatting, some Gaussians turn to become long and thin, generating artifacts when rendered into novel poses. We show the results without the scaling regularization in the middle column of Figure~\ref{fig:ablation}.

\subsection{Discussion}
\begin{figure}[ht]
    \includegraphics[width=\linewidth]{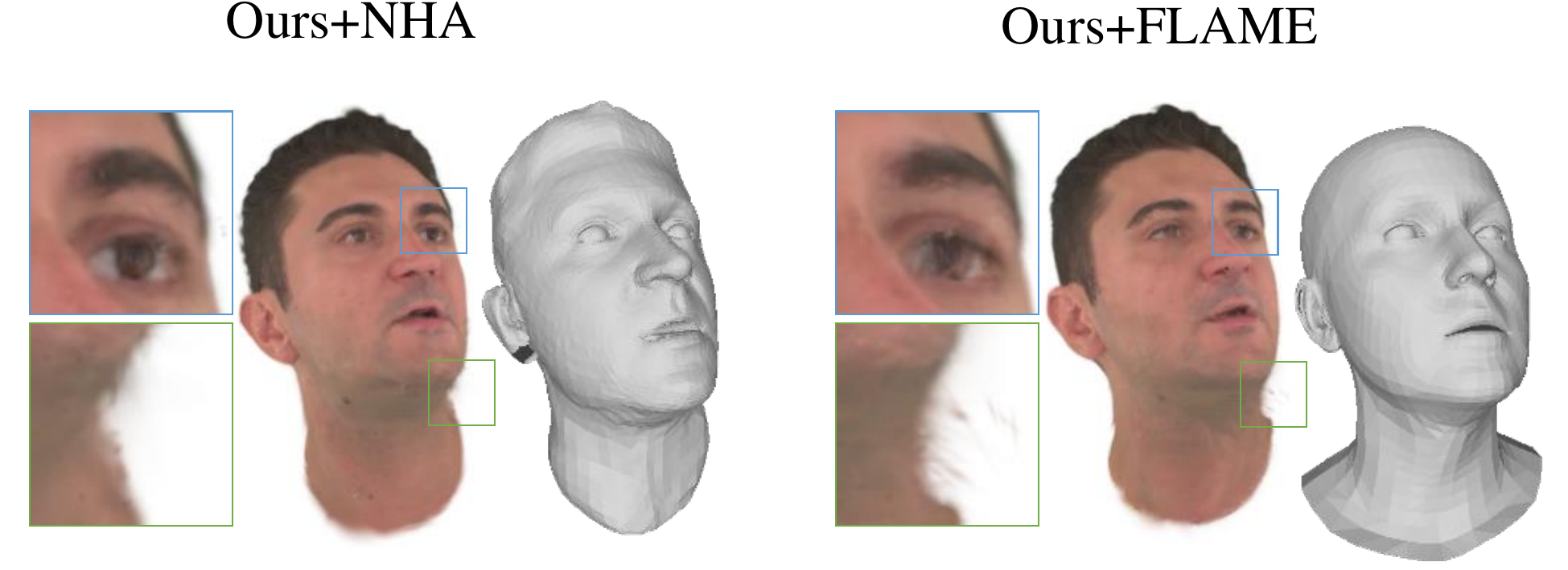}
    \caption{ 
        \textbf{Comparison between \emph{Ours+FLAME} and \emph{Ours+NHA}.} 
        The better aligned mesh from NHA~\cite{neural_head_avatars} improves the generalizability of SplattingAvatar to large pose variations.
    }
    \label{fig:ablation_mesh}
\end{figure}

\textbf{Discussion on driving mesh.}
Considering efficiency, compatibility, and portability, SplattingAvatar is designed to tightly rely on the motion and surface deformation of the underlying mesh. In the comparison between \emph{Ours+FLAME} and \emph{Ours+NHA}, we observe that the driving mesh should focus on the motion instead of fully reconstructing the exact geometry. 
In Figure~\ref{fig:ablation_mesh} we show that when the mesh with vertex offsets from NHA~\cite{neural_head_avatars} is applied, 
% the mesh naturally becomes the convex hull of the Gaussians, improving the generalizability of SplattingAvatar to large viewing angles.
the detailed surface deformation improves the generalizability of SplattingAvatar to large poses.
However, in the second and third row of Figure~\ref{fig:exp1}, the mesh from FLAME that captures the correct motion of the glasses rather than the shape is driving the best rendering quality of SplattingAvatar. 
To perform textured mesh rendering, the mesh of NHA~\cite{neural_head_avatars} is seamed in the mouth region and deformed to fit the shape of the glasses, yet both being unhelpful to the quality of \emph{Ours+NHA}.
% In the hair region, both methods perform similarly
% we do not observe any significant difference between the results using NHA mesh and FLAME.

\noindent\textbf{Limitations and future work.}
As discussed above, our method depends on the motion representation ability of the driving mesh. With current FLAME and SMPL-X models, we do not have separate motion representations for clothes and hair. We believe SplattingAvatar can support future works on human avatars with disentangled mesh representations, e.g., separate meshes for clothes and hair stands.

% \begin{table}[t]
% \centering
% \resizebox{\columnwidth}{!}{%
% \begin{tabular}{lccc}
%     \toprule
%      & PSNR$\uparrow$ & SSIM$\uparrow$ & LPIPS$\downarrow$ \\
%     \midrule
%     $\Delta x$, w/o trainable embedding & 16.00 & 0.897 & 0.246 \\
%     ours, w/ trainable embedding & \textbf{33.01} & \textbf{0.982} & \textbf{0.020} \\
%     \bottomrule
% \end{tabular}
% }
% \caption{
%     \textbf{Ablation on trainable embedding.}
%     We compare our trainable embedding with fixed embedding and trainable local shift $\Delta x$ on PeopleSnapshot \emph{male-3-casual}.
% }
% \vspace{-0.05in}
% \label{tab:abalation_nouvd}
% \end{table}

% \subsection{Implementations.}

% We use $\lambda_{l} = 0.01$ and $\lambda_{s} = 1.0$ all through the experiments. 

% \subsection{Refine SMPL-X parameters.}

%% file: 10_conclusion.tex
\section{Conclusion}
\label{sec:conclusion}
In this paper, we have proposed a hybrid representation for human avatar modeling featuring Gaussian Splatting with trainable embeddings on a mesh. We extend lifted optimization to simultaneously optimize the parameters of the Gaussians and their embeddings.
Our method leverages the advantages of the explicit motion representation with a mesh and implicit rendering capability of Gaussian Splatting.
Compared with SoTA methods, our approach achieves the best rendering quality for both head and full-body avatars reconstructed from monocular videos and runs at real-time frame rates on a mobile device.
% Our method lays a foundation for future work with more disentangled meshes for motion control and unified rendering with Gaussian Splatting.
Our method lays a foundation for future work in Gaussian Splatting manipulation with mesh-based motion control.

%% file: 12_appendix.tex
\renewcommand{\thefigure}{A\arabic{figure}}
\setcounter{figure}{0}
\renewcommand{\thetable}{A\arabic{table}}
\setcounter{table}{0}

\twocolumn[{
\begin{center}
    \Large
    \textbf{\thetitle}\\
    \vspace{0.5em}Supplementary Material \\
    \vspace{1.0em}
    
    \captionsetup{type=figure}
    \includegraphics[width=\textwidth]{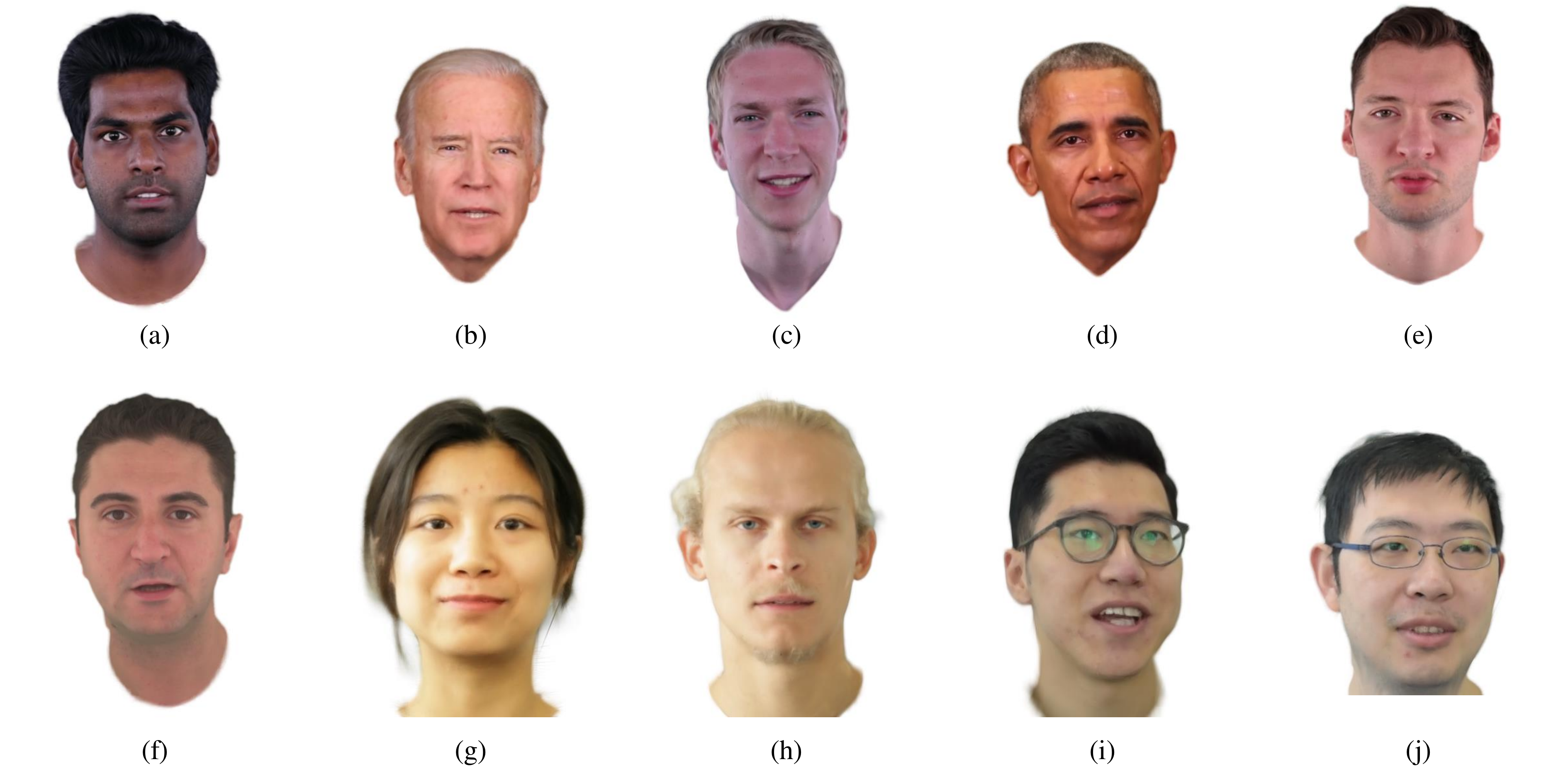}
    \captionof{figure}{
    \textbf{Dataset for head avatar.}
    We collected 10 subjects from publicly available datasets for the evaluation of head avatar modeling, with (a--e) from INSTA~\cite{INSTA}, (f) from NHA~\cite{neural_head_avatars}, (g, h) from IMAvatar~\cite{IMavatar:CVPR:2022}, and (i, j) from NerFace~\cite{NerFace:CVPR:2021}.
    We show the rendering results on the testing samples. 
    Our method captures high quality details, for example the light in the eyes, the texture of the hair, and off-surface geometry like the glasses.
    }
    \label{figa:suppl_dataset}
\end{center}
}]

\begin{figure*}[ht]
    \centering
    \includegraphics[width=\linewidth]{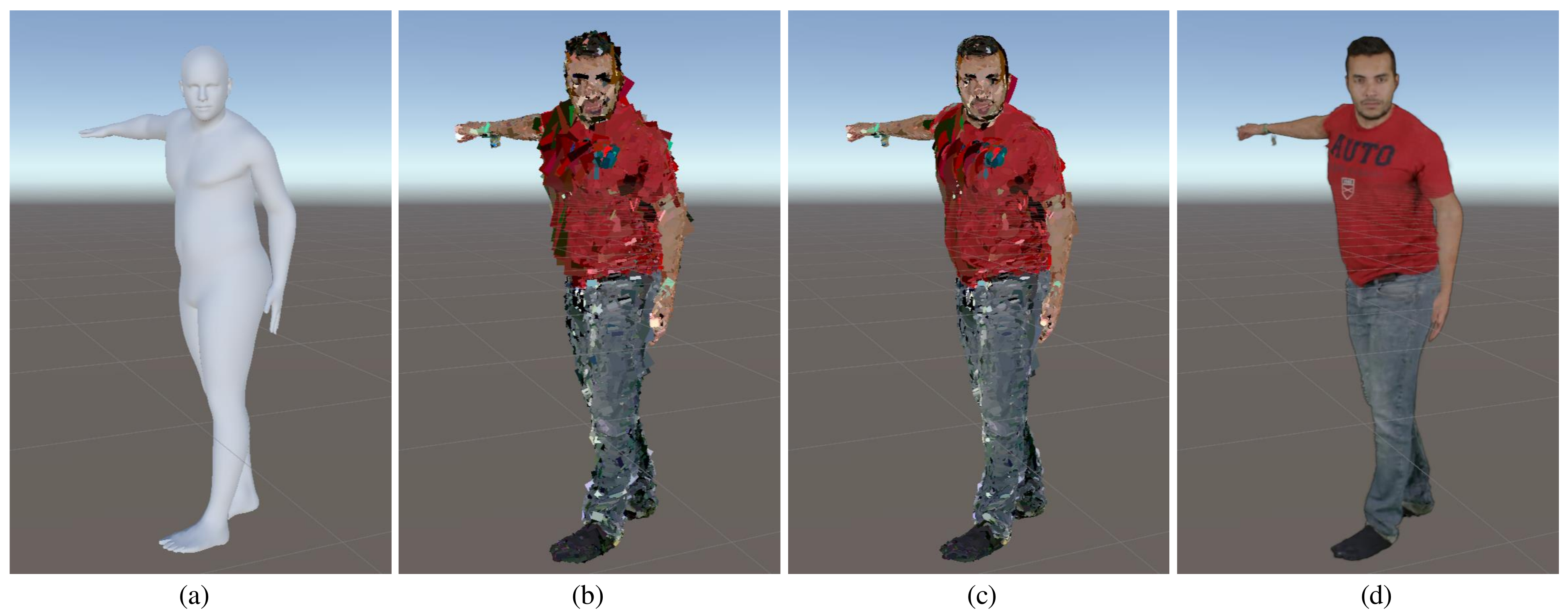}
    \caption{ 
        \textbf{Gaussian Splatting rendering in Unity.} 
        Our Unity implementation of Gaussian Splatting is conducted by drawing one quad primitive for each Gaussian. We show (a) the driving mesh for the current pose, (b) the quad primitive for each Gaussian, (c) the 2D covariance of the Gaussians illustrated by eclipses, and finally (d) the rendering result with \emph{$\alpha$-blending}.
    }
% \textcolor{ForestGreen}{green} 
% \textcolor{RubineRed}{red}
    \label{figa:suppl_unity}
\end{figure*}

\begin{figure}[ht]
    \includegraphics[width=\linewidth]{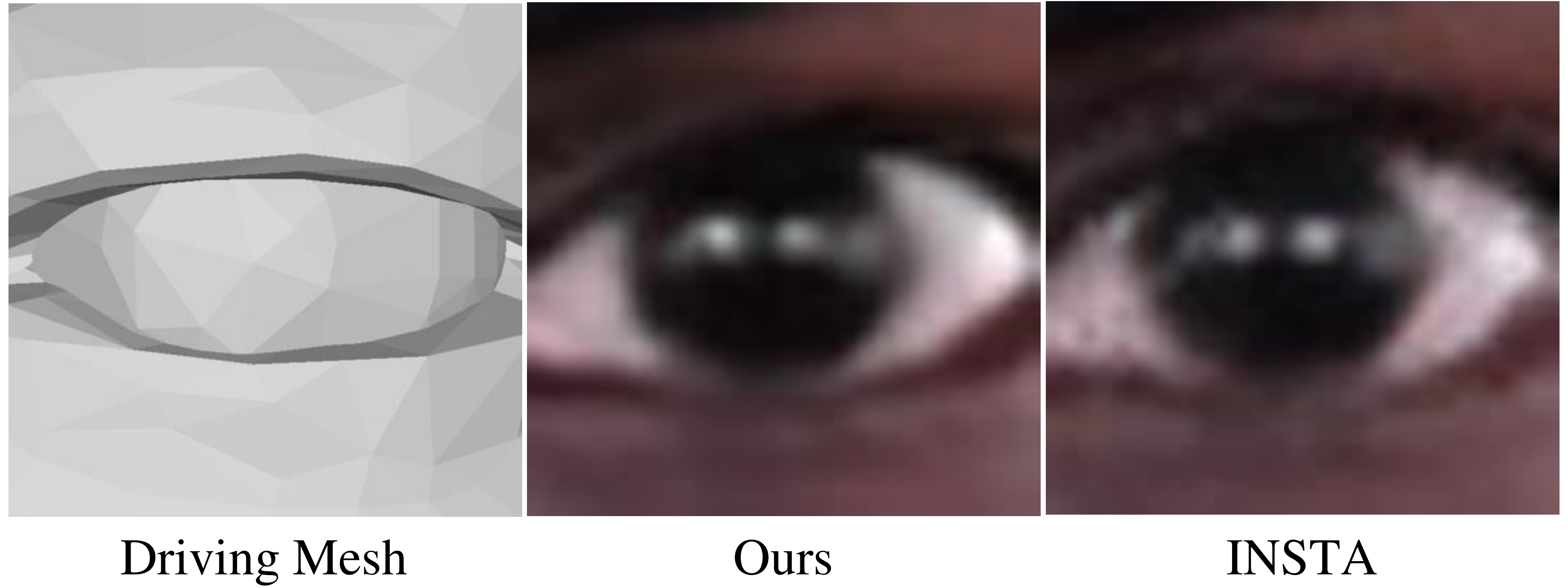}
    \caption{ 
        % \textbf{Effect of scaling regularization.} 
        % Without the scaling regularization term, Gaussians that are long and thin cause visible artifacts. Our scaling term successfully removes most of the artifacts when animated by novel poses.
        \textbf{Comparison with INSTA in the eye region.} 
        INSTA~\cite{INSTA} propose to find the nearest triangle when deforming a point in the posed space to the canonical space, causing unstable sampling in the canonical space and strong noise when dealing with complex geometries like the eye.
        Our embeddings-based motion control of the Gaussians leads to smooth rendering results.
    }
    \label{figa:suppl_eye}
\end{figure}

\begin{figure}[ht]
    \includegraphics[width=\linewidth]{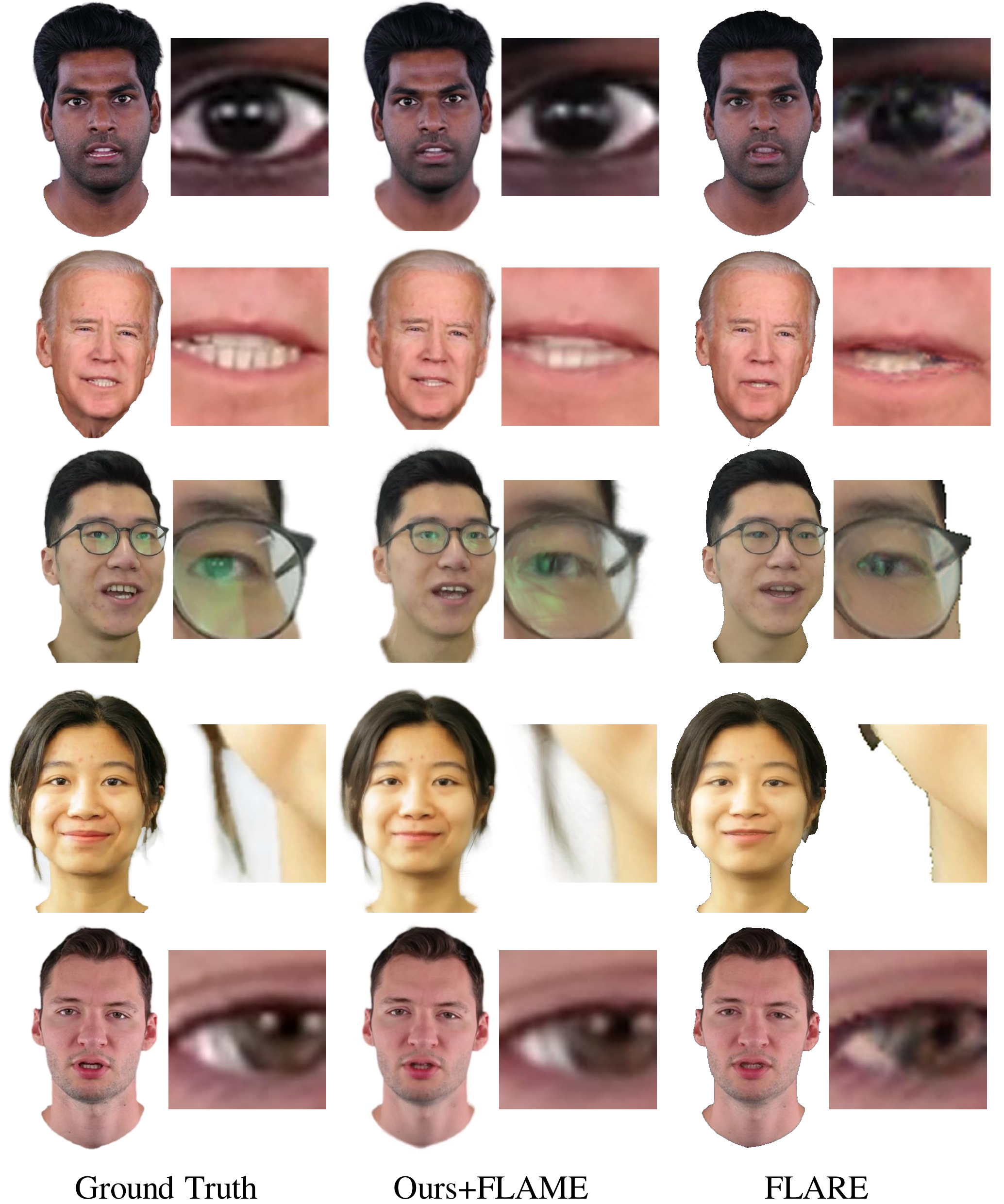}
    \caption{ 
        % \textbf{Effect of scaling regularization.} 
        % Without the scaling regularization term, Gaussians that are long and thin cause visible artifacts. Our scaling term successfully removes most of the artifacts when animated by novel poses.
        \textbf{Comparison with FLARE.} 
        We show qualitative comparison with FLARE~\cite{FLARE:SiggraphAsia:2023}.
    }
    \label{figa:suppl_exp1}
\end{figure}

\begin{figure*}[ht]
    \centering
    \includegraphics[width=\linewidth]{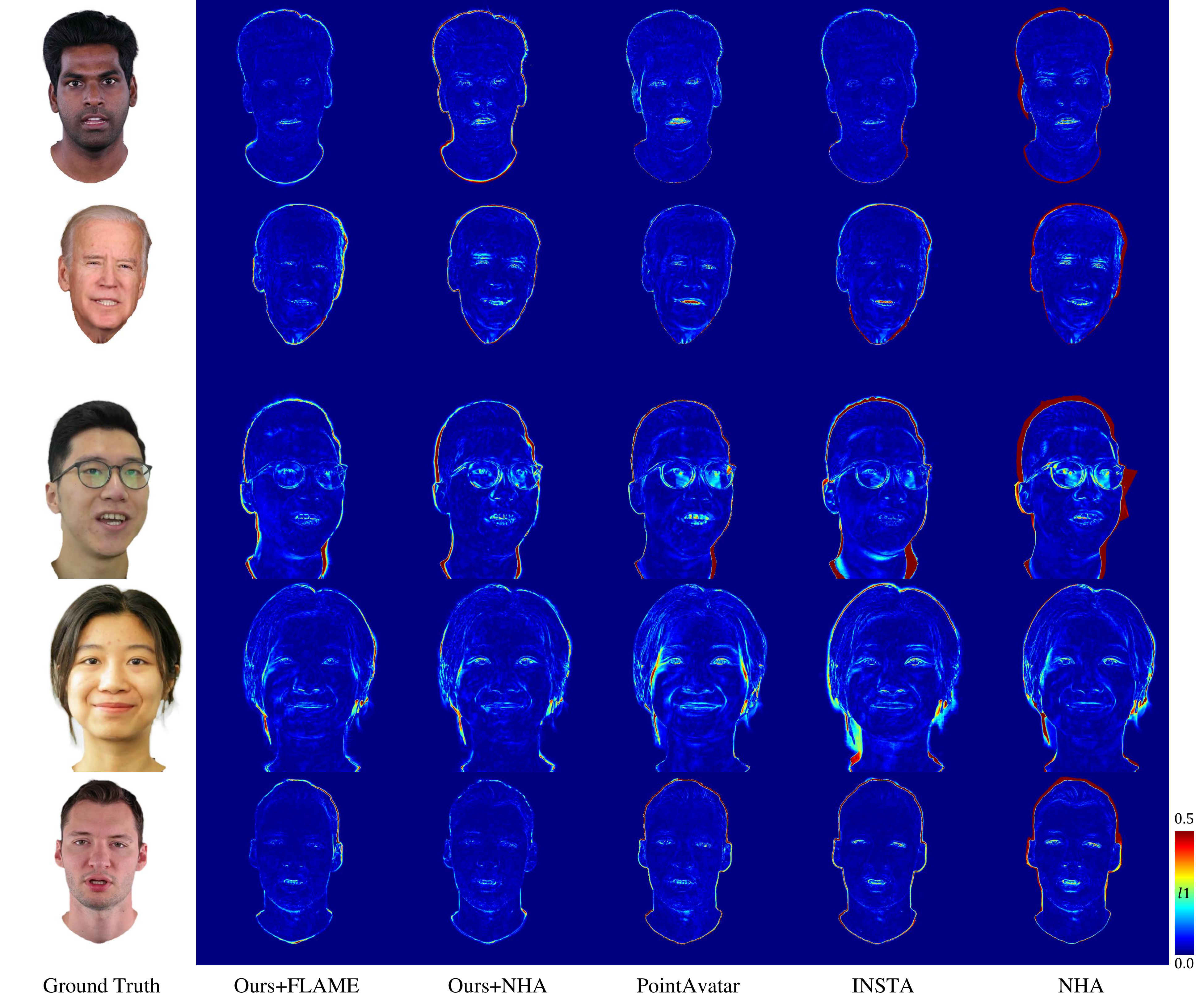}
    \caption{ 
        \textbf{Heatmaps of $l1$ error.} 
        We show the heatmaps illustrating the $l1$ RGB distance of the rendered images. Our methods and INSTA~\cite{INSTA} show overall better quality. The rendering quality of PointAvatar~\cite{PointAvatar} and NHA~\cite{neural_head_avatars} are limited by their point-based and mesh-based representations respectively.
    }
% \textcolor{ForestGreen}{green} 
% \textcolor{RubineRed}{red}
    \label{figa:suppl_error}
\end{figure*}

In this supplemental document, we elaborate 
details about the dataset for head avatar in Sec.~\ref{sec:suppl_dataset},
implementation details in Sec.~\ref{sec:suppl_impl}, 
and additional experimental comparisons in Sec.~\ref{sec:suppl_rlt}.

\section{Dataset}\label{sec:suppl_dataset}
In Figure~\ref{figa:suppl_dataset}, we show the 10 evaluated subjects that we collected from publicly available datasets, i.e., INSTA~\cite{INSTA}, NHA~\cite{neural_head_avatars}, IMAvatar~\cite{IMavatar:CVPR:2022}, and NerFace~\cite{NerFace:CVPR:2021}.
The rendering results are from \emph{Ours+FLAME}. Our method show high quality rendering capability with high fidelity details especially in the eyes, hair, and glasses.

\section{Implementation Details}\label{sec:suppl_impl}

\noindent\textbf{Training.}
We chose $\lambda_l = 0.01$, $\lambda_s = 1.0$, $T_s = 10.0$ and $T_r = 0.008$ all through the experiments.
We followed the original implementation of 3D Gaussian Splatting~\cite{kerbl3Dgaussians:SIGGRAPH:2023} to set the total number of iterations to 30,000 for each subject. 
Starting from iteration 600, the densify and prune process were conducted every 100 iterations. 
Every 3000 iterations, the opacity of all the Gaussians were reset to zero. We find this opacity-reset step effective in removing redundant Gaussians.
The densify, prune, and opacity-reset process stop at iteration 15,000.

\noindent\textbf{Unity rendering.}
As described in the main paper, in our Unity implementation, we draw one quad primitive for each Gaussian. The quad primitives are illustrated in Figure~\ref{figa:suppl_unity}. 
Benefiting from our trainable embedding scheme, the embeddings of the Gaussians were efficiently ported to compute shaders for the motion control of the Gaussians, leading to an animatable avatar running over 300 FPS on an NVIDIA RTX 3090 GPU.

\noindent
\textbf{Running time}.
With our pybind11 implementation, the \emph{walking on triangle} step takes around 3.5 ms. We conduct this step after densifying and pruning.
For comparison, \emph{densify-clone} takes 2.5 ms and \emph{densify-split} takes 6 ms.

The whole optimization follows the conversion of the original Gaussian Splatting that the number of total iterations is 30000, and the \emph{densify}, \emph{prune}, and \emph{walking on triangle} steps are performed every 100 iterations.

\section{Additional Results}\label{sec:suppl_rlt}

\noindent\textbf{Comparison with FLARE.}
FLARE~\cite{FLARE:SiggraphAsia:2023} is a mesh-based avatar modeling approach focusing on relightable avatar reconstructed from monocular videos, which is published very recently.
In Table~\ref{taba:suppl_flare}, we show comparison with FLARE on our head avatar dataset. FLARE~\cite{FLARE:SiggraphAsia:2023} reconstruct accurate geometry and materials of the avatar that our method does not focus on, while the strength of our method is the significant improvement in photometric quality and efficiency in rendering. Qualitative comparison is shown in Figure~\ref{figa:suppl_exp1}.

\noindent\textbf{Non-ambiguous motion control.}
One of the key benefits of our method is the non-ambiguous motion control comparing to the backward tracing process of NeRF-based avatar rendering.
INSTA~\cite{INSTA} propose to simplify this step by finding the nearest triangle for the deformation from the posed space to the canonical space.
We show in Figure~\ref{figa:suppl_eye} that this simplification causes significantly more noise when dealing with complex geometries like in the eye region.

\begin{table}[ht]
\centering
% \resizebox{\columnwidth}{!}{%
\begin{tabular}{lccc}
    \toprule
    Method & PSNR$\uparrow$ & SSIM$\uparrow$ & LPIPS$\downarrow$ \\
    \midrule
    FLARE~\cite{FLARE:SiggraphAsia:2023} & 23.87 & 0.893 & 0.129 \\
    \midrule
    Ours+FLAME & \underline{28.19} & \textbf{0.931} & \underline{0.063} \\
    Ours+NHA & \textbf{28.86} & \textbf{0.931} & \textbf{0.060} \\
    \bottomrule
\end{tabular}
% }
\caption{
    \textbf{Quantitative comparison with FLARE.}
    We show comparison with the recently published avatar modeling method FLARE~\cite{FLARE:SiggraphAsia:2023} on our head avatar dataset.
}
\vspace{-0.05in}
\label{taba:suppl_flare}
\end{table}

\noindent\textbf{Error map.}
Due to the limitation of segmentation and head tracking in the pre-processing pipeline. 
The metrics of photometric error in the main paper was affected by the error mostly in the neck area.
We show in Figure~\ref{figa:suppl_error} the error maps of the evaluated methods.
Our methods and INSTA~\cite{INSTA} show overall better quality. 
PointAvatar~\cite{PointAvatar} and NHA~\cite{neural_head_avatars} both focus on relightable modeling with explicit shape representations, which compromise their performance in terms of pixel-wise metrics.

\begin{table}[t]
\centering
% \footnotesize
% \resizebox{\columnwidth}{!}{%
\begin{tabular}{llccc}
    \toprule
    & Method & PSNR$\uparrow$ & SSIM$\uparrow$ & LPIPS$\downarrow$ \\
    \midrule
    \multirow{ 2}{*}{bala} & w/o walking & 29.91 & 0.933 & 0.070 \\
    & w/ walking & \textbf{30.04} & \textbf{0.938} & \textbf{0.062} \\
    \midrule
    \multirow{ 2}{*}{male-3-casual} & w/o walking & 32.48 & 0.979 & 0.024 \\
    & w/ walking & \textbf{33.01} & \textbf{0.982} & \textbf{0.020} \\
    \bottomrule
\end{tabular}
% }
\caption{
    \textbf{Quantitative ablation on \emph{walking on triangle}.}
    % The performance drop when turning off the \emph{walking on triangle} mechanism and clip UV values.
}
% \vspace{-0.15in}
\label{taba:abalation_walk_bala}
\end{table}
%\vspace{-0.20in}

\begin{figure}
    \centering
    \includegraphics[width=\linewidth]{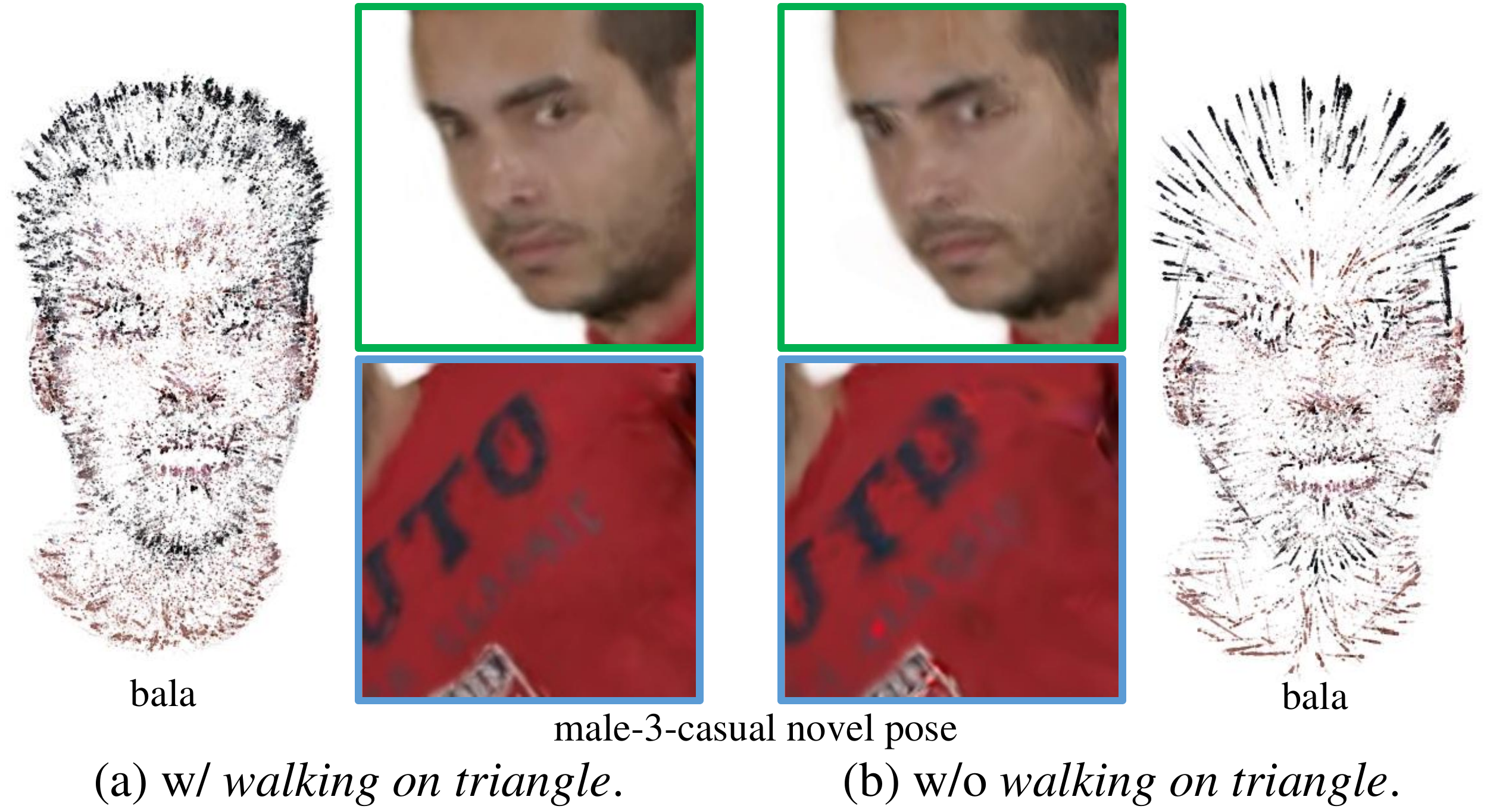}
    \caption{
        \textbf{Ablation on \emph{walking on triangle}.}
        Disabling \emph{walking on triangle} leads the Gaussians to stick and pile up on triangle boundaries, and cause artifacts when animated by novel poses.
    }
    \label{figa:ablation_walk}
\end{figure}

\noindent
\textbf{Ablation on \emph{walking on triangle}}.
We firstly conducted an ablation study on head avatar \emph{bala} where we disabled the \emph{walking on triangle} mechanism and clipped the UV values to prevent the Gaussians from moving beyond their corresponding triangles.
In addition to the performance drop as listed in Table~\ref{taba:abalation_walk_bala}, the Gaussians tend to stick and pile up on the boundaries of the mesh triangles as shown in Figure~\ref{figa:ablation_walk}.
The performance drop was more significant in the second experiment on full-body avatar \emph{male-3-casual}. Especially when animated by novel poses, turning off \emph{walking-on-triangle} resulted in noticeable artifacts.